\documentclass[12pt]{article}
\usepackage{graphicx}
\usepackage{epsfig}
\usepackage{color}
\usepackage{cite}

\begin{document}

\title{ The drastic outcomes from voting alliances in three-party bottom-up democratic voting\\ (1990 $\rightarrow$ 2013)}
\author{Serge Galam\\ Centre National de la Recherche Scientifique (CNRS), France}

\date{(serge.galam@cnrs-bellevue.fr)}

\maketitle

\begin{abstract}
The drastic effect of local alliances in three-party competition is investigated in democratic hierarchical bottom-up voting. The results are obtained analytically using a model which extends a sociophysics frame introduced in 1986  \cite{psy} and 1990  \cite{lebo} to study two-party systems and the spontaneous formation of democratic dictatorship. It is worth stressing that the 1990 paper was published in the Journal of Statistical Physics, the first paper of its kind in the journal. It was shown how a minority in power can preserve its leadership using bottom-up democratic elections. However such a bias holds only down to some critical value of minimum support. The results were used latter to explain the sudden collapse of European communist parties in the nineties. The extension to three-party competition reveals the mechanisms by which a very small minority party can get a substantial representation at higher levels of the hierarchy when the other two competing parties are big. Additional surprising results are obtained, which enlighten the complexity of three-party democratic bottom-up voting. In particular, the unexpected outcomes of local voting alliances are singled out.  Unbalanced democratic situations are exhibited with strong asymmetries between the actual bottom support of a party and its associated share of power at the top leadership. Subtle strategies are identified for a party to maximize its hold on the top leadership. The results are also valid to describe opinion dynamics with three competing opinions.

\end{abstract}


\section{Back to the future}

\subsection{A bit of history}

In the year 1990 the Journal of Statistical Physics has published  a paper, ``strange" at that time, named ``Social Paradoxes of majority Rules Voting and renormalization Group" in its Volume 61 \cite{lebo}. Before the paper went to print the Editor in Chief Joel Lebowitz wrote a handwritten letter to the author of the paper to notice him his personal negative view about the subject of the paper and its eventual potential for further research. However, contrary to the practice of some Editors, he followed, against his personal view, the recommendation to publish the paper from the Associate Editor Dietrich Stauffer who was handling the referee process. 

Such a personal rigor to keep up with the Journal procedures against a subjective feelings must be saluted. It is hard today to appreciate the full significance of such a choice due to the current radically different frame of mind with respect to interdisciplinary studies and the numerous applications of physics outside physics.

At those times, while I was advocating the use of modern Statistical Physics to describe social and political phenomena { \it \`a la mani\`ere} of an activist, all physicists besides a very few individuals, leading and non leading, old and young, were adamantly hostile to the very idea of Sociophysics. It was perceived as heretic, a total blaspheme,  a non sense, a wrong idea. Indeed to evoke the hypothesis that Òhumans could behave, even  in part, as atomsÓ was look upon as an evident absurdity. Note that outside physics the rejection was as total as inside physics, in particular among social scientists. A short testimony is given in \cite{testi} and a full report in \cite{book}.

However, despite the overwhelming ÒcondemnationÓ of the physicists's community I kept on my activism convinced of the rightness of my vision. The courageous attitude of Lebowitz to publish one of my first sociophysics paper combined with the energetic enthusiasm of the young Dietrich Stauffer, have helped me a good deal to keep me on the building of sociophysics. Those earlier papers, the epistemolgical and political ones,  which did envision the bursting of sociophysics are indeed totally unknown and I take the advantage of this special issue to cite them  \cite{1,2,3,4,5,6,7,8,9}.

It is worth to stress that to the best of my knowledge this publication \cite{lebo} has made the Journal of Statistical Physics the first physics international research journal ever to print a full paper whose content is the use of physics to describe a socio-political phenomenon. This earlier event grounds back the current special issue in the solid tradition of the editorial policy conducted by the Chief Editor Lebowitz. Today, sociophysics is an established stream of research in physics \cite{stauffer-book, inde,  fortunato-review, galam-review, book} with numerous topics of social sciences being investigated. In particular a good deal of papers has been devoted to the study of opinion dynamics \cite{sznajd,  redner-2, frank-voter,  espagnol,  herrmann, schneider, hetero, sousa, lambiotte-ausloos,  pierluigi, mariage,  martins1, martins2, contra, unify, lehir, inflex}.

\subsection{From two to three-party competition}

In the next Section the use of bottom-up democratic voting in hierarchical structures is reviewed  \cite{psy, lebo}. Natural and reasonable biases are shown to produce a major and unexpected breaking of the democratic balance expected from the use of majority rule, driving the stable establishment of democratic dictatorship. This demonstration is performed in the case of two competing parties A and B as it is often the case with all kind of competitions which end up as a binary race. Those effects were a result of the existence of ties at voting in even-size groups.

While the tie effect could have been perceived as univocally linked to the frame of bimodal dynamics, the extension to three-party competition was shown to preserve the feature with the appearance of another type of tie linked to the possible absence of majority in even-size groups with equal votes for each party. Such a robustness of the dictatorship effects was already discussed in those earlier works on bottom-up hierarchical voting but only schematically at specific cases \cite{psy, lebo, syst}. The tie resolution is no longer grounded on the necessity to have at least fifty percent plus one vote to win an election. It is a result of  local alliances among the various parties in competition.

It took fourteen years before  I completed a systematic investigation with Gelke and Peliti \cite{three} and only recently additional  results  were obtained  on lobbying effect \cite{book}. In parallel several  works have extended my earlier  investigation of three-opinion dynamics. One study choses to select the opinion at a tie either randomly or picking it up from the nearest group \cite{yang}. Others works consider the voter model  \cite{red, mobilia1}. The competition between persuasiveness and inflexibility was also analyzed \cite{mobilia2}. The problem of emergence of extremism has been also addressed by introducing a middle discrete option in the so called CODA model \cite{martins3}.

The rest of the paper is organized as follows: Section 2 sets back the problem as investigated in 1990. Going through the bottom-up voting the procedure to predict the outcome of an election is reviewed. Some new insight is provided for the pair case which has not bee treated in the earlier papers  \cite{psy, lebo, syst}. The emblematic case of groups of size 4 is described at some length. The recipe for building up the perfect democratic dictatorship is outlined in SubSection 2.2. The shift from from 2 to 3 parties  is studied in Section 3 where the first SubSection is devoted to the systematic determination of the bottom-up voting flow given all possible sets of local and global alliances among the three competing parties. The two dimensional landscape is found to include nine fixed points whose stability is studied. Section 4 illustrates a few meaningful cases of tricky dynamics produced by different sets of local alliances among the 3 parties. The case of a small party facing two opposite big ones is analyzed in details exhibiting subtle and counter-intuitive outcomes. The opposite case of a big party just below fifty percent facing two small ones both around twenty-five percent is also investigated. Last but not least, the voting flows for a set of four different bottom supports are exhibited each with eight different sets of alliances in a series of Figures.  Some concluding remarks end the paper.

\section{Setting back the problem as in 1990: dictatorship effect of two-party democratic competition}

The problem of bottom-up democratic voting in hierarchical structures is reviewed as studied in the earlier works of 1986  \cite{psy} and 1990  \cite{lebo}. Some new results are presented for the case of voting groups of size 2. 

Two parties A and B are competing to run a bottom-up democratic hierarchical structure. It may be a political group, a firm, a society or any human organization. Each member of the corresponding frame supports a specific party. At the moment of the vote for building up the hierarchy, the overall respective proportions in favor of A and B are $p_0$ and $(1-p_0)$ with the social entity concerned with the election.

The bottom level of the hierarchy is constituted from people who have been selected at random from the associated population. They are randomly organized in small groups of size $r$. Then, in a second step, each one of these $r$-sized groups elects a representative according to a local majority rule. Following the local majority in favor of either A or B the associated representative is a supporter of A or B.

For odd sized groups a local majority always exists for a two party competition with every member holding a specific support. However for even-size groups a tie may occur, resulting in a no majority situation. In these cases, mechanisms of different kinds can be evoked to justify that an A supporter is elected with a probability of $k$ and a B supporter with a probability $(1-k)$. The value of $k$ is a direct function of both the psycho-sociological nature and the history of the population as well as the current voting procedures.

Those elected agents constitute the first hierarchical level of the hierarchy denoted by level 1 with respect to the bottom level denoted 0. The local group voting process used at the bottom level is then repeated again identically to build a second level with now the voting groups formed by random selection of the elected representatives. Those groups elect their respective representative which in turn constitute the second level of the hierarchy.

\begin{figure}[ h]
\centering
\includegraphics[width=.8\textwidth]{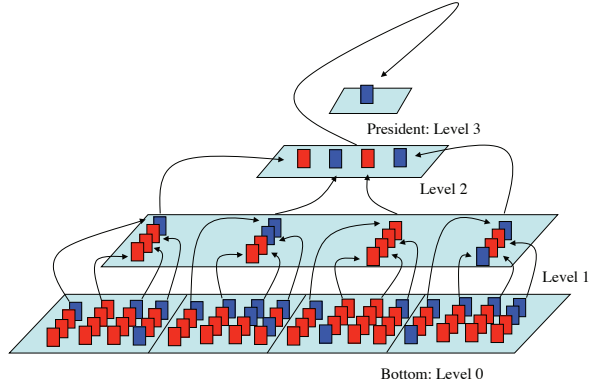}
\caption{A bottom-up hierarchy with 3 levels and voting groups of size 4.}
\label{h1}
\end{figure}

The process is repeated $n$ times using the same voting scheme to move from a level $l$ to the above level $(l+1)$. Each time, starting from the elected people at a given level, new groups are formed, which in turn elect new representatives to build up a higher level. At the top level is the president. However, since each one of these votes reduces the number of associated representatives by a factor $r$ from level $l$ to level $(l+1)$, it is necessary to have selected enough local bottom groups in order to ensure an $n$-level hierarchy. Therefore $r^n$ local groups are required at the bottom level 0 to have a succession of $n$ levels. This yields respectively $r^{n-1}$ elected representatives at level 1, $r^{n-2}$ elected representatives at level 2, $r^{n-l}$ elected representatives at level l and $r^{n-n}=1$ elected representative at level $n$. Figure (\ref{h1}) shows a bottom-up hierarchy with 3 levels and voting groups of size 4.

\subsection{Predicting the election outcome}

For group size $r$ the probability $p_{l+1}$ to have an A supporter elected at level $(l+1)$ can be calculated explicitly from $p_{l}$, the proportion of A supporter representatives at level $l$. The voting function $p_{l+1}=P_r(p_{l})$ is,

\begin{equation}
 P_r(p_l) =\sum_{m= \frac{r+1}{2} }^{r}  {r \choose m} p_l^m  (1-p_l)^{r-m} ,
\label{pr-odd} 
\end{equation}
for odd sizes $r$, and
\begin{equation}
 P_r(p_l)\equiv \sum_{m= \frac{r}{2} +1}^{r}  {r \choose m} p_l^m  (1-p_l)^{r-m}
+k {r \choose \frac{r}{2}} p_l^\frac{r}{2}  (1-p_l)^\frac{r}{2},
\label{pr-even} 
\end{equation}
when $r$ is even.

A broken democratic balance is monitored by $k$, in favor of A when $k>\frac{1}{2}$ and B for $k<\frac{1}{2}$. The democratic balance is restored only at $k=\frac{1}{2}$. Depending on the current values of $p_0$, $r$ and $n$, we have either $0<p_n<1)$ yielding a probabilistic outcome or $p_n\approx 0, 1$ allowing a deterministic result within some digit precision since bottom proportions are given by real numbers.  In the following probabilities are rounded to full percents with two digits after the decimal points unless otherwise stated.

In real systems, voting groups can be of different sizes making necessary to consider combination of those sizes pondered with their respective proportions making the voting function more complicated to solve but preserving the main features of the associated voting dynamics. Solving $P_r(p_l)=p_l$ the two fixed points $p_B=0$ and $p_A=1$ are always found. They are invariant with size change.

\begin{figure}[ h ]
\centering
\includegraphics[width=.5\textwidth]{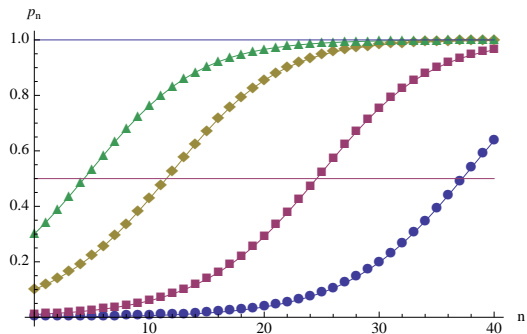}
\caption{Any small proportion $p_0$ grows while climbing up the hierarchy to eventually reach 1 provided the number of levels $n$ is sufficiently large. The cases $p_0=0.001, 0.01 , 0.10, 0.30$ are shown with $k=0.6$ for $r=2$.}
\label{pair1}
\end{figure}

For $r=2$ Eq. (\ref{pr-even}) writes,
\begin{equation}  
p_1\equiv P_2(p_0)=p_0^2+2 k  p_0 (1-p_0) \ , 
\end{equation}
making either $p_B=0$ or $p_A=1$ an attractor of the dynamics with the other one being unstable. Indeed, when $k>\frac{1}{2}$, any $p_0\neq 0$ would grow while climbing up the hierarchy to eventually reach the value of 1 provided the number of levels $n$ is sufficiently large, as seen in Figure (\ref{pair1}) for the cases $p_0=0.001, 0.01 , 0.10, 0.30$   with $k=0.6$.

\begin{figure}[ htpb ]
\centering
\includegraphics[width=.5\textwidth]{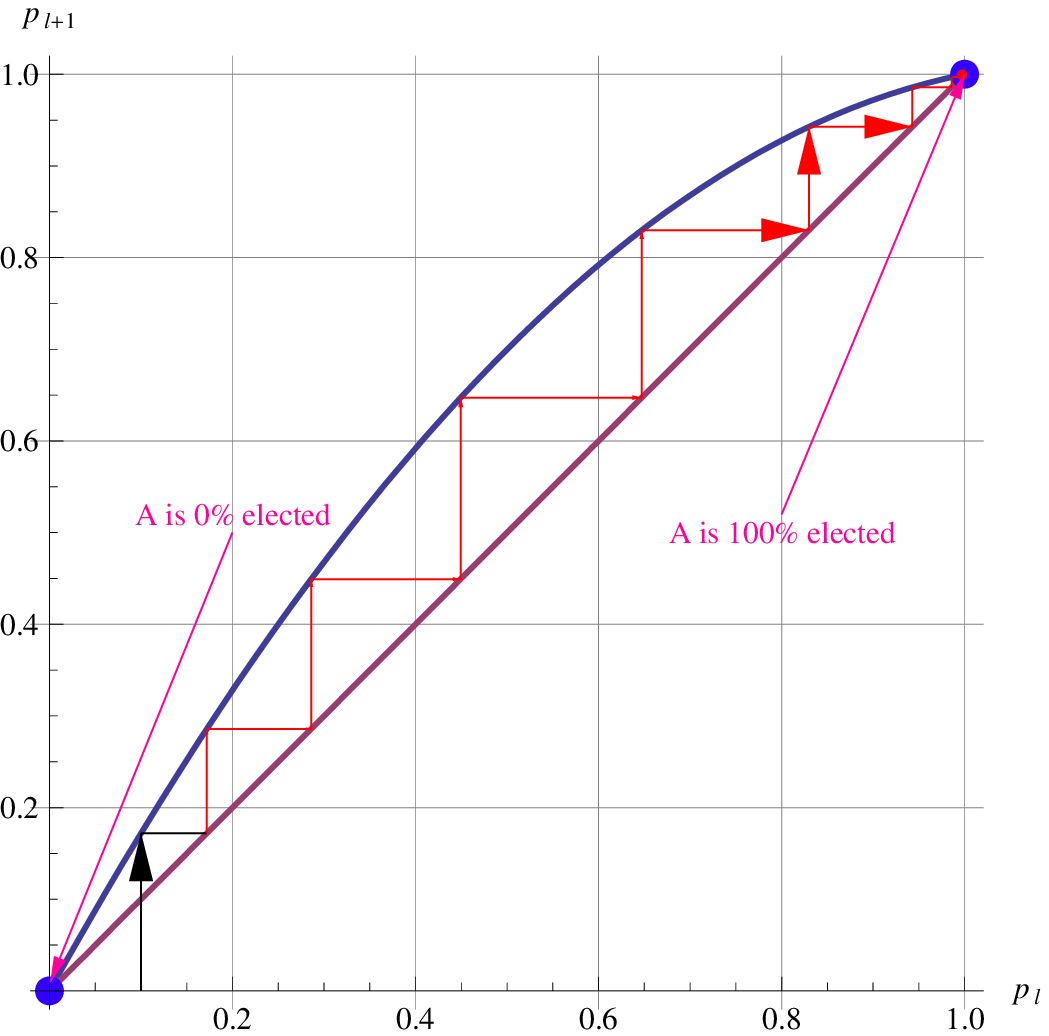}
\caption{The variation of $p_{l+1}$ as a function of $p_{l+1}$ for  $r=2$ at $k=0.9$. The evolution from $p_0=0.10$ is shown from roght to left. Eight levels are required to reach 1.}
\label{pair-plus}
\end{figure}

Even from a totally negligible $p_0=0.001$ support at the bottom the A party wins a good chance to get the presidency. For instance, 40 levels boost  $p_0=0.001$ up to $p_{40}=0.64$ while $p_0=0.01 , 0.10, 0.30$ yields respectively $p_{40}=0.96, 1 , 1$. However, no real hierarchy will have forty levels, most having much less, usually of the order 10.  Yet , $p_0=0.10$ and 0.30 yields  $p_{10}=0.85, 0.97$ respectively giving a significant chance to win the presidency. The dynamics of $p_{l+1}$ as a function of $p_{l+1}$ is illustrated in Figure (\ref{pair-plus}) for the case $k=0.9$ together with the evolution from $p_0=0.10$  shown from right to left for eight levels.

Such a radical effect in favor of the status quo of pair interactions has been substantiated in some nice work by Krzysztof Kulakowski and Maria Nawojczyk \cite{mariage} on the votes of married couples in the United States. 

Contrary to the very unbalanced precedent $r=2$ case the next smallest size of $r=3$ ensures a democratically balance with
\begin{equation}  
p_{l+1}\equiv P_3(p_l)=p_l^3+3 p_l^2 (1-p_l) \ .
\label{p1} 
\end{equation}
Besides $p_B=0$ and $p_A=1$ a third unstable fixed point  $p_{c,3}=\frac{1}{2}$ appears. Now, repeating votes drive towards either full power at $p_A$ for $p_0>\frac{1}{2}$ or total disappearance at $p_B$ when $p_0<\frac{1}{2}$. The dynamics is democratic since the bottom majority secures the top outcome.

For instance $p_0=0.45$ yields the series $p_1=0.42$, $p_2=0.39$, $p_3=0.34$, $p_4=0.26$, $p_5=0.17$, $p_6=0.08$ down to $p_7=0.02$ before $p_8=0.00$. A strong minority of $45\%$ is self-eliminated within 8 levels. Fewer levels give a chance to the minority of running the organization. The situation is perfectly symmetrical with respect to A and B. Both share the same threshold to full power. 

The bottom-up voting structure simultaneously allocates local power to the minority party and ensures the organization presidency to the majority opinion given enough numbers of levels. It costs very little in terms of manpower since only a small part of the associated population is needed to implement and activate the hierarchy. Above dynamics is invariant for an increase in the size $r$ as long as odd values are used. Only the number of levels decreases and the number of people increases.

At $r=4$, the next up size, three fixed points are obtained as before with still $p_B=0$ and $p_A=1$ but the unstable fixed point $p_{c,r}$ is no longer at $\frac{1}{2}$ turning the dynamics non democratic.  The associated threshold value becomes asymmetric for rulers and non rulers. For $r=4$ the probability to have a representative A elected given by Eq. (\ref{pr-even}) is,
\begin{equation}  
p_{l+1}\equiv P_4(p_l)=p_l^4+4 p_l^3 (1-p_l)+6 k p_l^2 (1-p_l)^2 \ ,  
\end{equation}
where the last term embodies the bias in favor of A. For the extreme case $k=0$ the tipping point for A is thus,
\begin{equation} 
p_{c,4}=\frac{1+\sqrt{13}}{6} \approx 0.77 \ ,  
\end{equation}
which is drastically shifted from $\frac{1}{2}$. Simultaneously, the B threshold to stick to power is down to $\approx 23\%$. Eq. (6) is independent of $n$ and determines the flux of the probabilities going up the hierarchy for A with $p_0>p_{c,4} \rightarrow p_{l+1}>p_l$ against $p_0<p_{c,4} \rightarrow p_{l+1}<p_l$. Simultaneously the tipping point for B to increase or decrease its representation is 
$(1-p_{c,4})$. The number of levels fixes the final probability for the presidency. When it is sufficiently large the probability becomes either one or zero.

To stick to power, provided there exists sufficient levels, the B needs only to get over $23\%$ of the bottom support while to take over  power, the A needs to boost their support at above $77\%$. The bottom-up voting is no longer democratic. 

In addition the number of levels to democratic self-elimination is even smaller than in the democratic $r=3$ case. For $r=4$ and $k=0$, the above $r=3$ series changes to $p_0=0.45$, $p_1=0.24$, $p_2=0.05$ and $p_3=0.00$. Instead of 8 levels, 3 are enough to make the A disappear from the elected representatives. Even $p_0=0.70$ yields $p_1=0.66$, $p_2=0.57 $, $p_3=0.42 $, $p_4=0.20 $, $p_5=0.03 $, and $p_6=0.00$ making a support of $70\%$ self-eliminated democratically after six bottom-up levels.  

An a priori reasonable bias in favor of the ruling party with $k=1$ motivated by the statement ``to make a change you need a majority",   turns a majority rule democratic voting into a totalitarian outcome. 

Increasing the size of even groups weakens the breaking of the democratic symmetry between the two parties as expected since increasing the size makes the chance of getting a tie less probable. From $p_{c,4}=\frac{1+\sqrt{13}}{6}$ for size 4 we have $p_{c,r}=\frac{1}{2}$ for $ r \rightarrow \infty$. Larger voting group sizes also reduce the number of levels necessary to get to the stable fixed points, which in turn guarantee a deterministic result.

\begin{figure}[ h]
\centering
\includegraphics[width=.5\textwidth]{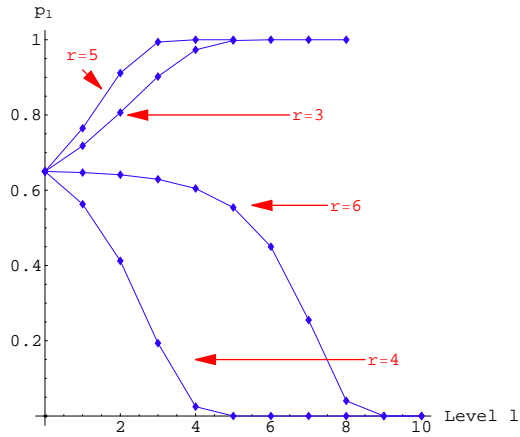}
\caption{ Climbing up the hierarchy $p_0=0.65$ as a function of the hierarchical level for voting group sizes of $r=3,4,5,6$. The cases $r=3$ and $r=5$ leads to A ruling from the fifth level for $r=3$ and the fourth level for $r=5$. The cases $r=4$ and $r=6$ leads to B ruling although the A party has a stronger majority at the bottom. However, the B party rules with certainty from the fifth level using $r=4$ and only from the ninth one for $r=6$. }
\label{pl-r}
\end{figure}

Figure (\ref{pl-r}) shows the results of repeating voting for $p_0=0.65$ using voting groups of respective sizes of $r=3,4,5,6$. The bottom-up voting outcomes are shown as a function of the hierarchical level. 

\subsection{Building the perfect democratic dictatorship}

Above results show that to start beyond the threshold guarantees an increasing weight in terms of the elected representatives. However to be certain to win at the presidency level, the minimum number of levels required is a function the bottom support $p_0$. Accordingly, to calculate the critical number of levels which ensures running the organization, provided that  $p_0>p_{c,r}$, is obtained by expanding the voting function $p_l=P_r(p_{l-1})$ around the unstable fixed point $p_{c,r}$ with,
\begin{equation}  
p_l\approx p_{c,r}+(p_{l-1}-p_{c,r}) \lambda_r \ , 
\label{nc1} 
\end{equation}
where $\lambda_r \equiv \frac{dP_r(p_n)}{dp_n}|_{p_{c,r}}$ and  $P_r(p_{c,r})=p_{c,r}$. Rewriting Eq. (\ref{nc1}) as
\begin{equation}  
p_l-p_{c,r}\approx (p_{l-1}-p_{c,r}) \lambda_r \ ,  
\end{equation}
and iterating from level $l$ down to level $0$ we get,  
\begin{equation}  
p_l-p_{c,r}\approx (p_0-p_{c,r}) \lambda_r^l \ ,  
\end{equation}
which yields,
\begin{equation}  
p_l\approx p_{c,r}+(p_0-p_{c,r}) \lambda_r^l \ ,
\label{nc1-1}
\end{equation}
which in turn leads to,
\begin{equation}  
l \approx \frac{1}{\ln \lambda_r}  \ln\left(\frac{ p_l-p_{c,r}}{ p_0-p_{c,r}} \right) \ .
\label{nc2} 
\end{equation}

Writing $\Delta_{top}\equiv p_{top}-p_{c,r}$ and $\Delta_{bot}\equiv p_{0}-p_{c,r}$ Eq. (\ref{nc2}) becomes
\begin{equation}  
n_{c,r} \approx \frac{1}{\ln \lambda_r}  \ln\left(\frac{\Delta_{top}}{\Delta_{bot}}\right)  \ ,
\label{nc3} 
\end{equation}
where $n_{c,r}$ is the critical number of levels requires to reach the value $p_{top}$ at the top of the hierarchy starting from $p_0$ at the bottom. Similarly  $\Delta_{top}$ and $\Delta_{bot}$ measure the distance to the tipping point  from respectively the top and the bottom of the hierarchy. 

\begin{figure}[ h]
\centering
\includegraphics[width=.5\textwidth]{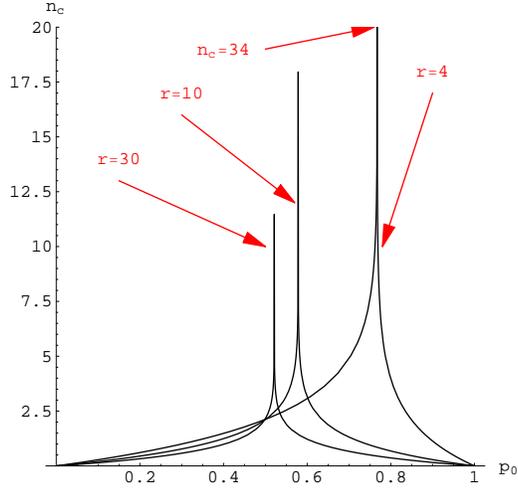}
\caption{The variation of the critical number of levels $n_{c,r}$ as a function of $p_0$ for three group sizes $r=4, 10, 30$.}
\label{ncr}
\end{figure}

Eq. (\ref{nc3}) yields two different critical numbers of levels $n_{c,r}^+$ ($p_0>p_{c,r}$, $\Delta_{top}=1-p_{c,r}$) and $n_{c,r}^-$ ($\Delta_{top}=-p_{c,r}$ for $p_0<p_{c,r}$) for the respective certainty to win and loose the presidency for A. The variation of  $n_{c,r}$ as a function of $p_0$ for  group sizes $r=4, 10, 30$ is exhibited in Figure (\ref{ncr}).

Although above expansions are expected to be valid only for $p_0$ in the vicinity of $p_{c,r}$ they are found to be rather good estimates over the whole of the range of values from 0 to 1. This is not the case for Eq. (\ref{nc1-1}). 

Rewriting Eq. (\ref{nc1-1}) to extract the value of $p_0$ we have,
\begin{equation}  
p_0=p_{c,r}+(p_l-p_{c,r}) \lambda_r^{-l} \ .  
\label{pcn}
\end{equation}
Therefore, given an $n$-level hierarchy using voting groups of size $r$ the critical value of bottom support $p_{0,B}^{r,n}$ at and below which the B party wins the presidency is given by
\begin{equation} 
p_{0,B}^{r,n}=p_{c,r}(1-\lambda_r^{-n}) \ ,
\label{pcn1}
\end{equation}
having plugged $p_l=0$, which means that at level $l$ the A party is certain to get zero elected representative.

At the other extreme $p_n=1$ gives the bottom threshold $p_{0,A}^{r,n}$ from which the challenging A party wins the presidency. We thus get
\begin{equation}  
p_{0,A}^{r,n}=p_{0,B}^{r,n}+\lambda_r^{-n}\ ,
\label{pcn2}
\end{equation}
where $p_{0,B}^{r,n}$ is given by Eq. (\ref{pcn1}).

Three regimes are obtained, of which two are deterministic and one is probabilistic. For $p_0<p_{0,B}^{r,n}$, $p_m=0, \forall m\geq n$ while $p_m=1, \forall m\geq n$ when $p_0>p_{0,a}^{r,n}$. In between with $p_{0,B}^{r,n}<p_0<p_{0,A}^{r,n}$ we have $0<p_n<1$, i.e., the vote outcome for presidency is probabilistic. This intermediate probabilistic regime determines a coexistence window where some democracy is prevailing since no party is sure of winning, which makes the shift of leadership a reality. However Eq. (\ref{pcn2}) shows that this democratic window shrinks as a power law $\lambda_r^{-n}$ of the number $n$ of hierarchical levels. We have $p_{0,A}^{r,n}\Rightarrow p_{0,B}^{r,n}$ where $n>>1$.

\section{From two up to three competing parties}

Above study of two competing parties relies on the broken tie hypothesis, which indeed required the existence of even-size voting groups. However, tie breaking may also occur for odd sizes when three parties A, B, and C are competing with A B C configurations. 
There, the tie cannot be resolved by giving an advantage to the ruling party. While for two parties it makes sense that the challenging party must have at least half plus one votes to win the election, for three parties at a tie each party has one third of the votes, which is less than fifty of the votes, so none is legitimated to have an advantage. Therefore, an agreement among two parties at the expense of the third one is necessary to win the election. 

Before starting one can be tempted to ask why not to solve an apparently simpler case considering groups of size 2? First reason is because size 2 exhibits a tie already with two opinions belonging to the even size family. The second cause is the fact that  3 opinions produce 3 tie configurations A B, A C and C D requesting three independent parameters to define what happens for each case. Therefore,  the problem takes place in a three dimensional space. At the contrary, for size 3, while no tie exists for two opinions,  for three opinions there exists only one tie configuration A B C with two independent parameters to define what happens as it is shown below. Therefore size 3 is left for future work.

\begin{figure}[ htpt]
\centering
\includegraphics[width=1\textwidth]{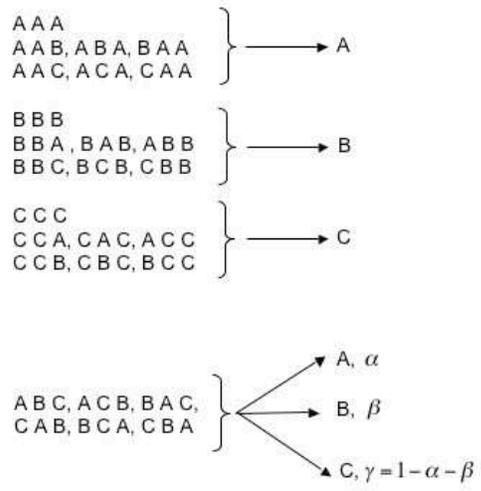}
\caption{The voting rules for groups of size 3.}
\label{abc}
\end{figure}

Voting rules are shown in Figure (\ref{abc}). Local majority wins making 2 A, 2 B, 2 C to elect respectively an A, B, or C, whatever the third agent. All six configurations with 1 A, 1 B, 1 C require a voting agreement between two parties to elect a representative. Party alliances can be global or local. In this case, to account for all possibilities we introduce a probability $\alpha$ to have an A elected, a probability $\beta$ to have a B elected and a probability $\gamma$ to have a C elected under the constraint $\alpha +\beta + \gamma =1$. For the sake of the presentation we select $\alpha$ and $\beta$ as the independent parameters with $\gamma =1-\alpha -\beta$.

\begin{figure}
\centering
\includegraphics[width=1.5\textwidth]{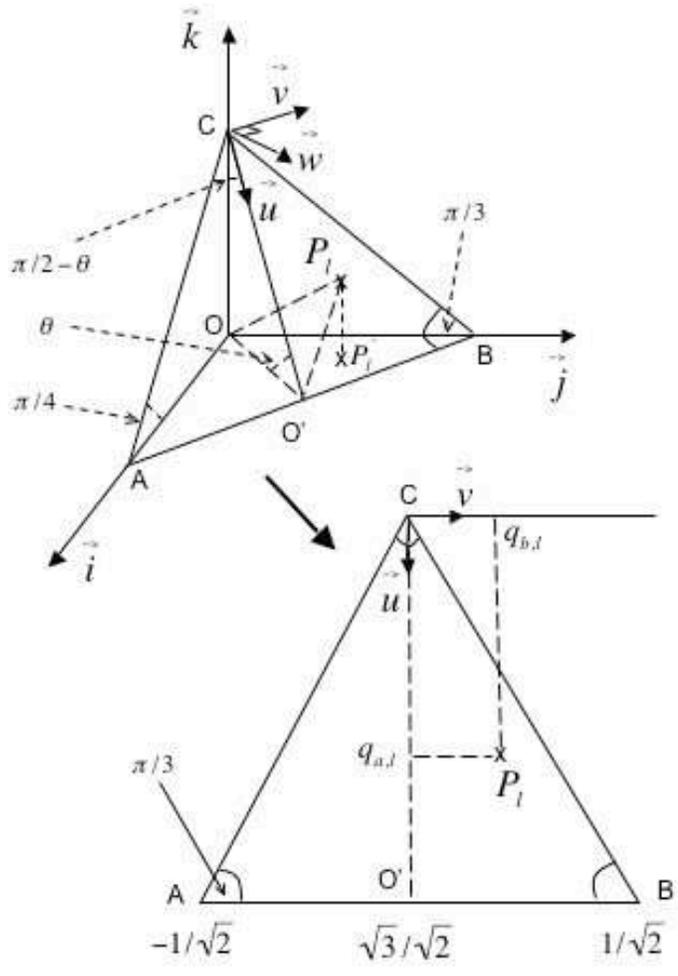}
\caption{The three party problem reduces to a two-dimensional problem.}
\label{xyz0}
\end{figure}

Denoting $p_{a,l}, p_{b,l}, p_{c,l}$ of A, B and C representatives the respective proportions of elected representatives of party A, B and C at level $l$, we have $p_{a,l+1}, p_{b,l+1}, p_{c,l+1}$  at level $(l+1)$. These proportions obey the constraint $p_{a,l}+ p_{b,l}+ p_{c,l}=1$ since we assume that every agent supports one party. The associated voting functions for respectively A, B and C are,
\begin{equation}  
p_{a,l+1}\equiv P_{a,3}(p_{a,l},p_{b,l})=p_{a,l}^3+3 p_{a,l}^2 (1-p_{a,l}) +6 \alpha p_{a,l} p_{b,l} (1-p_{a,l}- p_{b,l})\ ,  
\label{abc-a}
\end{equation}
\begin{equation}  
p_{b,l+1}\equiv P_{b,3}(p_{a,l},p_{b,l})=p_{b,l}^3+3 p_{b,l}^2 (1-p_{b,l}) +6 \beta p_{a,l} p_{b,l} (1-p_{a,l}- p_{b,l}) \ ,  
\label{abc-b}
\end{equation}
\begin{equation}  
p_{c,l+1}=(1-p_{a,l+1}- p_{b,l+1})\ ,  
\label{abc-c}
\end{equation}
where Eq. (\ref{abc-c}) reduces the problem to two dimensions with $(p_{a,l}, p_{b,l}, p_{c,l})$ confined onto the triangle shown in Figure (\ref{xyz0}). A point $P_l$ associated to  $p_{a,l}, p_{b,l}, p_{c,l}$ is represented simultaneously in both coordinate systems as
\begin{equation}  
\vec{OP_l}\equiv  p_{a,l} \vec{i} +  p_{b,l} \vec{j}+  p_{c,l} \vec{k}=\vec{OC}+q_{a,l} \vec{u} +  q_{b,l} \vec{v}+  q_{c,l} \vec{w} ,
\label{xyz1}
\end{equation}
with $(p_{a,l}, p_{b,l}, p_{c,l})$ for  $(O, \vec{i}, \vec{j}, \vec{k})$ and $(q_{a,l}, q_{b,l}, q_{c,l})$ for  $(C, \vec{u}, \vec{v},\vec{w})$.

Having (see Figure (\ref{xyz0}))
\begin{equation}
\left \{
\begin{array} {c}
\vec{u}= \frac{\sqrt{6}}{6} ( \vec{i} +  \vec{j}) - \frac{\sqrt{2}}{\sqrt{3}} \vec{k} , \\ \\
\vec{v}= \frac{\sqrt{2}}{2}(- \vec{i} +  \vec{j}) ,\\ \\
\vec{w}= \frac{\sqrt{3}}{3} (\vec{i} +  \vec{j}+  \vec{k}),
\end{array}
\right .
\label{xyz2}
\end{equation}
we get,
\begin{equation}
\left \{
\begin{array} {c}
q_{a,l}=   \frac{\sqrt{6}}{2}( p_{a,l}+p_{b,l}) ,\\ \\
q_{b,l}=  \frac{1}{\sqrt{2}}(-p_{a,l}+p_{b,l})   ,\\ \\
q_{c,l}= - 1 +p_{a,l}+p_{b,l}+p_{c,l}=0  .
\end{array}
\right .
\label{xyz4}
\end{equation}
Eqs. (\ref{xyz4}) demonstrates the reduction of the three dimensional space onto the two-dimensional triangle $ABC$ of Figure (\ref{xyz0}) spanned by $(C, \vec{u}, \vec{v})$. The coordinates (1, 0, 0), (0, 1, 0), (0, 0, 1) of $A, B, C$ become  $(\sqrt{3}/\sqrt{2}, - \sqrt{2}/2)$, $(\sqrt{3}/\sqrt{2}, \sqrt{2}/2)$ and $(0, 0)$. We thus have $-\sqrt{2}/2 \leq q_{a,l} \leq \sqrt{2}/2$ and $ 0 \leq q_{b,l} \leq \sqrt{3}/\sqrt{2}$ with the constraint $ 0\leq q_{a,l} +q_{b,l} \leq 1$.

\subsection{Determining the bottom-up voting flow}

Solving the set of fixed point Equations
\begin{equation}  
\left \{
\begin{array} {c}
p_{a,l+1}=p_{a,l} \ ,  \\æ\\
p_{b,l+1}=p_{b,l} \ ,  \\æ\\
p_{c,l+1}=p_{c,l}.
\end{array}
\right .
\label{abc-0}
\end{equation} 
allows to build the bottom-up voting phase diagram. Putting one proportion equal to zero makes the two other fixed point equations identical to the two party case. For instance $p_{c}^*=0$ satisfies $p_{c,l+1}=p_{c,l}$, which in turn reduces $p_{a,l+1}=p_{a,l}$ and $p_{b,l+1}=p_{b,l}$ to the form of Eq. (\ref{p1}) with $p_{l+1}==p_l^3+3 p_l^2 (1-p_l)$. 

\begin{figure}[ htpt]
\centering
\includegraphics[width=1.5\textwidth]{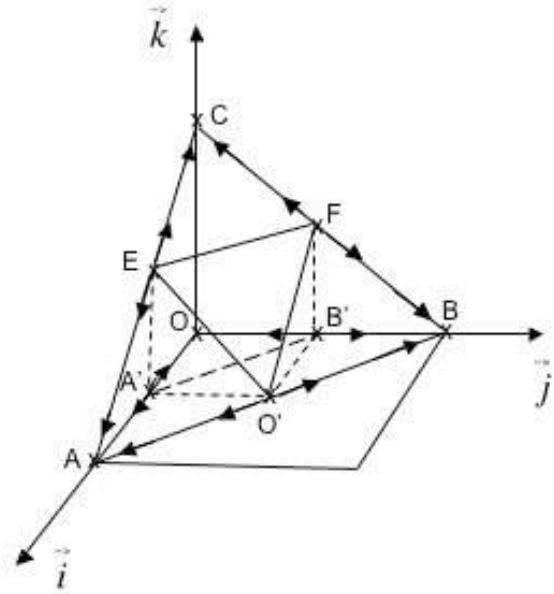}
\caption{The six fixed points A, B, C, E, F, O' are shown together with their projections A, B, O, A', B', O' onto the plane $(O, \vec{i}, \vec{j})$. The respective stability of each point is indicated by arrows.}
\label{fixed-pt3}
\end{figure}

We thus have $p_{c}^*=0$ with  $p_{a}^*=0, 1/2, 1$ and $p_{b}^*=0, 1/2, 1$. All associated combinations $(p_{a}^*, p_{b}^*, p_{c}^*)$ are valid solutions of Eq. (\ref{abc-0}) provided their sum is equal to one with then only the two solutions  $(0, 1, 0)$ and $(1/2, 1/2, 0)$. Permutations on a, b, c, yields  $(0, 1, 0)$, $(0, 0, 1)$, $(1/2, 0, 1/2)$, $(0, 1/2,1/2)$ amounting to a total of six fixed points, 
\begin{itemize}
\item A for $(1, 0, 0)$, 
\item B for $(0, 1, 0)$, 
\item C for $(0, 0, 1)$,
\item O' for $(1/2, 1/2, 0)$,
\item E for $(1/2, 0, 1/2)$,
\item F for $(0, 1/2, 1/2)$.
\end{itemize}
in $(O, \vec{i}, \vec{j}, \vec{k})$ and,
\begin{itemize}
\item A with $(1, 0, 0)$, 
\item B with $(0, 1, 0)$, 
\item O with $(0, 0, 0)$,
\item O' with $(1/2, 1/2, 0)$,
\item A' with $(1/2, 0, 0)$,
\item B' with $(0, 1/2, 0)$,
\end{itemize}
in $(O, \vec{i}, \vec{j})$ as shown in Figure (\ref{fixed-pt3}) where the associated flows to each one of the two party combinations along the three one dimensional lines A-O'-B, A-E-C, B, F, C together with the projections onto the $(O, \vec{i},\vec{j})$ are included.

The six fixed points A, B, C, O', E, F as well as their projection in the $(O, \vec{i}, \vec{j})$ plane are independent of the value of $\alpha$ and $\beta$ since indeed they correspond to the limit cases where one proportion is zero with A, E, C corresponding to $p_b=0$; B, F, C to $p_a=0$; and A, O', B to $p_c=0$. In such a case no tie occurs since the problem is reduced to 2 parties. The respective flows along each line AC, CB, BA, are also independent of $\alpha$ and $\beta$ for the same reason.

However, before searching for eventual additional fixed points, Eq. (\ref{abc-0}) has in principle nine solutions, it is worth first investigating the stability of all six fixed points A, B, C, O', E, F shown in Figure (\ref{fixed-pt3}). This will allow us to reduce the search for numerical solutions by determining the $(\alpha, \beta)$ range where extra fixed points are expected inside the triangle ABC.

Performing a Taylor expansion of Eqs. (\ref{abc-a}, \ref{abc-b}) around a fixed point $(p_{a}^*, p_{b}^*)$ with respect to $p_{a,l}$ and $p_{b,l}$ we get

\begin{equation}  
\left \{
\begin{array} {c}
d_{a,a}\equiv\frac{\partial P_{a,3}(p_{a,l},p_{b,l})}{\partial p_{a,l}}=-6p_{a,l}^2+6 p_{a,l}+6 \alpha p_{b,l} (1-2p_{a,l}-p_{b,l}) \ ,  \\æ\\
d_{a,b}\equiv\frac{\partial P_{a,3}(p_{a,l},p_{b,l})}{\partial p_{b,l}}=6 \alpha p_{a,l} (1-p_{a,l}-2p_{b,l}) \ ,  
\end{array}
\right . 
\label{ta1}
\end{equation}
and
\begin{equation}  
\left \{
\begin{array} {c}
d_{b,a}\equiv\frac{\partial P_{b,3}(p_{a,l},p_{b,l})}{\partial p_{a,l}}=6 \beta p_{b,l} (1-2p_{a,l}-p_{b,l}) \ ,  \\æ\\
d_{b,b}\equiv\frac{\partial P_{b,3}(p_{a,l},p_{b,l})}{\partial p_{b,l}}=-6p_{b,l}^2+6 p_{b,l}+6 \beta p_{a,l} (1-p_{a,l}-2p_{b,l}) \ .
\end{array}
\right . 
\label{tb1}
\end{equation}
Plugging  the coordinates $(p_{a}^*, p_{b}^*)$ in Eqs. (\ref{ta1}, \ref{tb1}) leads to, 
\begin{equation}  
\left (
\begin{array} {c}
p_{a,l+1}- p_{a}^*  \\æ\\
p_{b,l+1} - p_{b}^*   
\end{array}
\right )
=
M
\left (
\begin{array} {c}
p_{a,l}- p_{a}^*  \\æ\\
p_{b,l} - p_{b}^*   
\end{array}
\right )
\label{tab3}
\end{equation}
where 
\begin{equation}  
M=\left (
\begin{array} {c c}
d_{a,a}^* & d_{a,b}^*  \\æ\\
d_{b,a}^* & d_{b,b}^*
\end{array}
\right ) \ .
\label{m}
\end{equation}
We then diagonalize the matrix M to find out its associated eigenvalues
\begin{equation}  
M^*=\left (
\begin{array} {c c}
\lambda_{a}^* & 0  \\æ\\
0 & \lambda_{b}^*
\end{array}
\right ) \ .
\label{dm}
\end{equation}
and the associated two eigenvectors,
\begin{equation}  
V_{a}^*=\left (
\begin{array} {c}
V_{a,a}^*  \\æ\\
V_{a,b}^* \end{array}
\right ) \ ,
V_{b}^*=\left (
\begin{array} {c}
V_{b,a}^*  \\æ\\
V_{b,b}^* \end{array}
\right ) \ .
\label{vdm}
\end{equation}
Plugging in Eq. (\ref{m})  the A, B, C, O', E, F  six fixed point coordinates we get respectively,
\begin{equation}  
M_A=M_B=M_C=M_O=\left (
\begin{array} {c c}
0 & 0  \\æ\\
0 & 0
\end{array}
\right ) \ ,
\label{m1}
\end{equation}

\begin{equation}  
M_{O'}=\left (
\begin{array} {c c}
\frac{3}{2}(1-\alpha) & -\frac{3}{2}\alpha  \\æ\\
-\frac{3}{2}\beta & \frac{3}{2}(1-\beta)
\end{array}
\right ) \ ,
\label{m2}
\end{equation}

\begin{equation}  
M_E=M_{A'}=\left (
\begin{array} {c c}
\frac{3}{2} & \frac{3}{2}\alpha  \\æ\\
0 & \frac{3}{2}\beta
\end{array}
\right ) \ ,
\label{m3}
\end{equation}

\begin{equation}  
M_F=M_{B'}=\left (
\begin{array} {c c}
\frac{3}{2} \alpha & 0  \\æ\\
\frac{3}{2}\beta & \frac{3}{2}
\end{array}
\right ) \ .
\label{m4}
\end{equation}
which in turn yield the respective eigenvalues
\begin{equation}  
\left \{
\begin{array} {c}
\lambda_{a}^A=\lambda_{a}^B=\lambda_{a}^C=\lambda_{a}^O=0
\\æ\\
\lambda_{b}^A=\lambda_{b}^B=\lambda_{b}^C=\lambda_{b}^O=0
\end{array}
\right . \ ,
\label{dm1}
\end{equation}

\begin{equation}  
\left \{
\begin{array} {c}
\lambda_{a}^{O'}=\frac{3}{2} 
\\æ\\
\lambda_{b}^{O'}=\frac{3}{2}(1-\alpha-\beta)
\end{array}
\right . \ ,
\label{dm2}
\end{equation}

\begin{equation}  
\left \{
\begin{array} {c}
\lambda_{a}^{E}=\lambda_{a}^{A'}=\frac{3}{2} 
\\æ\\
\lambda_{b}^{E}=\lambda_{b}^{A'}=\frac{3}{2}\beta
\end{array}
\right . \ ,
\label{dm3}
\end{equation}

\begin{equation}  
\left \{
\begin{array} {c}
\lambda_{a}^{F}=\lambda_{a}^{B'}=\frac{3}{2} 
\\æ\\
\lambda_{b}^{F}=\lambda_{b}^{B'}=\frac{3}{2}\alpha
\end{array}
\right . \ .
\label{dm4}
\end{equation}
For the first series of the zero matrix Eq. (\ref{m1}) any vector is an eigenvector and in particular along  $\vec{i}$ and $\vec{j}$, which gives,

\begin{equation}  
V_{a,A}=V_{a,B}=V_{a,C}=V_{a,O}=\left (
\begin{array} {c}
1  \\æ\\
0\end{array}
\right ) \ ,
\label{m5}
\end{equation}

\begin{equation}  
V_{b,A}=V_{b,B}=V_{b,C}=V_{b,O}= \left (
\begin{array} {c}
0 \\æ\\
 1
\end{array}
\right ) \ .
\label{m5-1}
\end{equation}
For the other series, the associated eigenvectors are,
\begin{equation}  
V_{a,O'}=\left (
\begin{array} {c}
-1  \\æ\\
1\end{array}
\right ) \ ,
V_{b,O'}=\left (
\begin{array} {c}
\frac{\alpha}{\beta}  \\æ\\
 1
\end{array}
\right ) \ ,
\label{m6}
\end{equation}

\begin{equation}  
V_{a,E}=V_{a,A'}=\left (
\begin{array} {c}
1  \\æ\\
0\end{array}
\right ) \ ,
V_{b,E}=V_{b,A'}=\left (
\begin{array} {c}
\frac{-\alpha}{1-\beta}  \\æ\\
 1
\end{array}
\right ) \ ,
\label{m7}
\end{equation}

\begin{equation}  
V_{a,F}=V_{a,B'}=\left (
\begin{array} {c}
0 \\æ\\
1
\end{array}
\right ) \ ,
V_{b,F}=V_{b,B'}=\left (
\begin{array} {c}
-\frac{1-\alpha}{\beta}  \\æ\\
 1
\end{array}
\right ) \ .
\label{m8}
\end{equation}

Having $\lambda_{a}^A=\lambda_{a}^B=\lambda_{a}^C=\lambda_{a}^O=\lambda_{b}^A=\lambda_{b}^B=\lambda_{b}^C=\lambda_{b}^O=0$ indicates that the fixed points A, B, C are stable in all directions as well as O within the projection onto the $(O, \vec{i}, \vec{j})$ plane. On the contrary, the result $\lambda_{a}^{O'}=\lambda_{a}^{E}=\lambda_{a}^{A'}=\lambda_{a}^{F}=\lambda_{a}^{B'}=\frac{3}{2}$ makes the O', A', B' fixed points unstable along the $\vec{i} $ direction and O' E, F unstable along the transformed direction in the ABC plane obtained using the transformation Eq. (\ref{xyz4}). Indeed the corresponding eigenvectors in the plane $(C, \vec{u},\vec{v})$ are,

\begin{equation}  
W_{a,A}=W_{a,B}=W_{a,C}=W_{a,O}=\left (
\begin{array} {c}
\sqrt{\frac{3}{2}} \\æ\\
\frac{1}{\sqrt{2}}\end{array}
\right ) \ ,
\label{m9}
\end{equation}

\begin{equation}  
W_{b,A}=V_{b,B}=W_{b,C}=W_{b,O}= \left (
\begin{array} {c}
\sqrt{\frac{3}{2}} \\æ\\
-\frac{1}{\sqrt{2}}\end{array}
\right ) \ .
\label{m10}
\end{equation}
For the other series the associated eigenvectors are less ``sexy" in their expression with 
\begin{equation}  
W_{a,O'}=\left (
\begin{array} {c}
0
\\æ\\
\sqrt{2}
\end{array}
\right ) \ ,
W_{b,O'}=\left (
\begin{array} {c}
\sqrt{\frac{3}{2}} (\frac{\alpha +\beta}{\beta}) \\æ\\
(\frac{-\alpha +\beta}{\sqrt{2}\beta })
\end{array}
\right ) \ ,
\label{m11}
\end{equation}

\begin{equation}  
W_{a,E}=V_{1,A'}=\left (
\begin{array} {c}
\sqrt{\frac{3}{2}}  \\æ\\
-\frac{1}{\sqrt{2}}\end{array}
\right ) \ ,
W_{b,E}=V_{2,A'}=\left (
\begin{array} {c}
\sqrt{\frac{3}{2}} \frac{-1+\alpha+\beta}{-1+\beta}  \\æ\\
 \frac{-1-\alpha+\beta}{\sqrt{2}(-1+\beta)}
\end{array}
\right ) \ ,
\label{m12}
\end{equation}

\begin{equation}  
W_{a,F}=W_{1,B'}=\left (
\begin{array} {c}
\sqrt{\frac{3}{2}}  \\æ\\
\frac{1}{\sqrt{2}}
\end{array}
\right ) \ ,
W_{b,F}=W_{2,B'}=\left (
\begin{array} {c}
\frac{1}{\beta}\sqrt{\frac{3}{2}} (-1+\alpha+\beta)  \\æ\\
\frac{1-\alpha+\beta}{\sqrt{2}\beta}
\end{array}
\right ) \ .
\label{m13}
\end{equation}

All the first eigenvectors, which correspond to the 3/2 eigenvalue, are along the sides of the triangle ABC or its projection OAB as seen in Figure (\ref{fixed-pt3}). In addition, for each one of them, the stability along the other eigevector direction, i.e., the direction towards the interior of the triangle ABC, depends on $\alpha$ and $\beta$ as seen from their respective second eigenvalue $\lambda_{b}^{O'}, \lambda_{b}^{E}=\lambda_{b}^{A'}, \lambda_{b}^{F}=\lambda_{b}^{B'}$ (Eqs. (\ref{dm2}, \ref{dm3}, \ref{dm4})). We thus have:
\begin{itemize}
\item Stability for O', i.e., the flow goes towards the side of the triangle, when $\gamma=(1-\alpha-\beta)<\frac{2}{3}$. The direction is unstable otherwise and the flow goes towards the interior of the triangle when $(\alpha+\beta)>\frac{1}{3}$.
\item Stability for (E, A'), i.e., the flow goes towards the side of the triangle, when $\beta<\frac{2}{3}$. The direction is unstable otherwise and the flow goes towards the interior of the triangle when $\beta>\frac{2}{3}$. 
\item Stability for (F, B'), i.e., the flow goes towards the side of the triangle, when $\alpha<\frac{2}{3}$. The direction is unstable otherwise and the flow goes towards the interior of the triangle when $\alpha>\frac{2}{3}$. 
\end{itemize}

As can be seen from Figure (\ref{fixed-pt3}), the basin of attraction of each stable fixed points C, A, B and O, A, B include respectively the triangle CEF, AEO', BFO' and OA'B', AA'O', BB'O'. Depending on the value of the biases $\alpha$ and $\beta$ the strategic question is to determine which party benefits from the inside triangles A'B'O' and EFO'. In other words, which part of that triangle coalesces with the three incompressible basins of attraction?

From the above findings it is seen that if one of the flows, which goes towards the interior of the triangle, i.e., either $\alpha$, $\beta$ or $\gamma$ is larger than two thirds, then the two other values must be simultaneously lower than two thirds since $\alpha + \beta + \gamma=1$. This means that if one direction towards the inside of the triangle is unstable, both others are necessarily stable. However, while only one direction can be unstable at a time, the three directions can be simultaneously stable, i.e., $\alpha < 2/3$, $\beta < 2/3$ and $\gamma < 2/3$.

One straightforward way to resolve the question directly is to extract all possible nine solutions from Eq. (\ref{abc-0}). However the difficulty is that it is not possible to find all the solutions analytically. To be more precise, I was not able to do it, at least with compact and tractable expressions. Some intuitive and hand waving arguments are thus welcome to simplify the search for the eventual relevant solutions.

\section{Illustrating the tricky dynamics}

\subsection{A small party versus two large competing ones }

Many situations are characterized by the competition between two big parties A and B, of which none has the absolute majority. While they usually defend very different policies on key issues, there often exists a third party C, small in terms of support, whose emphasis is to ensure some specific interest of a given minority. In that conjecture one of the two large parties must get the support of C  to ensure the hierarchy presidency.

\begin{figure}[ h]
\centering
\includegraphics[width=.5\textwidth]{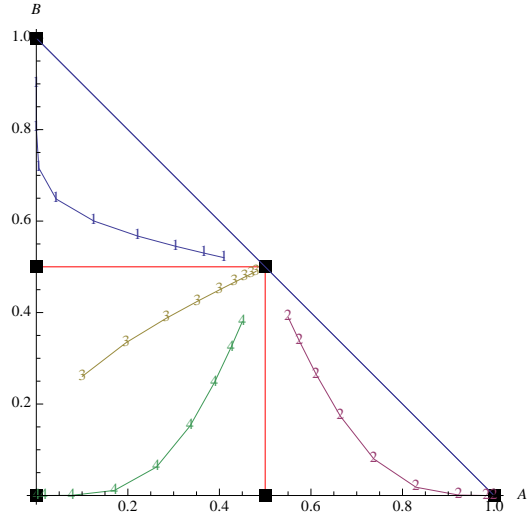}
\caption{The alliance set follows $(\alpha =0, \beta =0, \gamma =1 )$. As soon as $p_{a,0}<1/2$ and $p_{b,0}<1/2$, C will grow in power holding by climbing up the hierarchy. Four bottom supports are considered $p_{a,0}=0.41, p_{b,0}=0.52, p_{c,0}=0.17$ (noted with ``1");  $p_{a,0}=0.55, p_{b,0}=0.39, p_{c,0}=0.06$ (noted with ``2"); $p_{a,0}=0.48, p_{b,0}=0.49, p_{c,0}=0.03$ (noted with ``3"); $p_{a,0}=0.45, p_{b,0}=0.38, p_{c,0}=0.17$ (noted with ``4"). The C domain for power increase (a square) is double that  for A and B (isosceles triangle). The black squares are the various fixed points with the triangle vertices being the three attractors. }
\label{abc4}
\end{figure}

We consider first the case of a voting group size of 3 making the assumption that at any local tie, i.e., at a (A B C) configuration, it is a C agent who is elected whether a coalition is set up with A or B. The voting Equations are written as,
\begin{equation}  
p_{a,l+1}\equiv P_{a,3}(p_{a,l})=p_{a,l}^3+3 p_{a,l}^2 (1-p_{a,l}) \ ,  
\label{abc-a0}
\end{equation}
\begin{equation}  
p_{b,l+1}\equiv P_{b,3}(p_{b,l})=p_{b,l}^3+3 p_{b,l}^2 (1-p_{b,l})  \ ,  
\label{abc-b0}
\end{equation}
\begin{equation}  
p_{c,l+1}\equiv P_{c,3}(p_{c,l})=p_{c,l}^3+3 p_{c,l}^2 (1-p_{c,l}) +6 p_{a,l} p_{b,l} p_{c,l}\ ,  
\label{abc-c0}
\end{equation}
where Eqs. (\ref{abc-a}) and (\ref{abc-b}) are identical to the voting function for two competing parties. Therefore for A and B the critical threshold to win the hierarchy presidency requires a bottom support of above $50\%$. The corresponding number of levels to reach the presidency with certainty is also unchanged. Accordingly, when neither A nor B has more than fifty percent of support, the bottom-up voting increases the proportion of C representatives while decreasing those of A and B as seen from the flow diagram Figure (\ref{abc4}). The domain of attraction for the bottom small party C is twice as large as those of A and B.  

Four different bottom supports for A and B are shown  in Figure (\ref{abc4})  with respectively $p_{a,0}=0.41, p_{b,0}=0.52, p_{c,0}=0.17$  (noted with ``1");  $p_{a,0}=0.55, p_{b,0}=0.39, p_{c,0}=0.06$ (noted with ``2"); $p_{a,0}=0.48, p_{b,0}=0.49, p_{c,0}=0.03$ (noted with ``3"); $p_{a,0}=0.45, p_{b,0}=0.38, p_{c,0}=0.17$ (noted with ``4"). The black squares are the various fixed points. The triangle vertices are the three attractors. Only $p_{a,0}=0.55>1/2$ drives the A party towards the presidency as $p_{b,0}=0.52>1/2$ does for B. Any bottom support within the rectangle leads towards increasing C power.

\begin{figure}[ h]
\centering
\includegraphics[width=.5\textwidth]{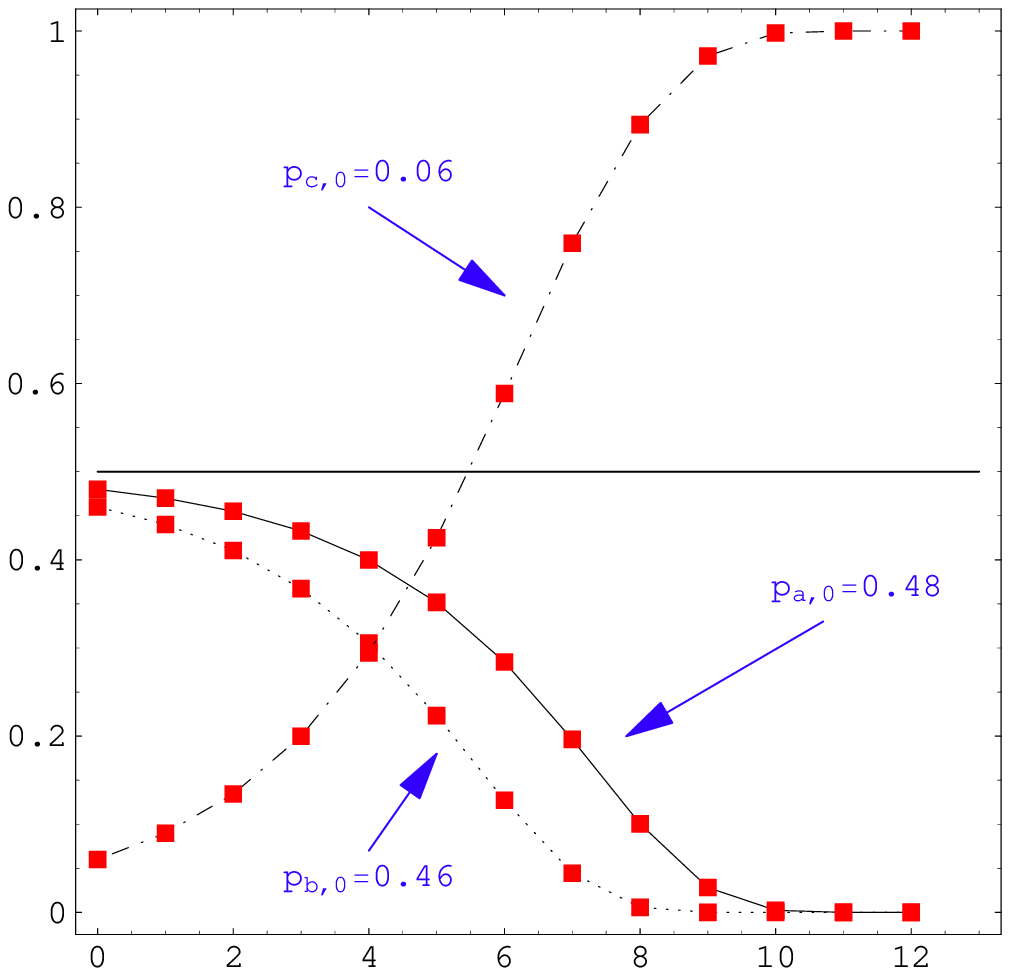}
\caption{The respective variation of power sharing for A, B, C from a bottom support $p_{a,0}=0.48$ (solid line), $p_{b,0}=0.46$ (dotted line), $p_{c,0}=0.06$ (dashed line) as a function of the hierarchy level up to level 12. The tiny minority party C starts from six percent of support at the bottom and yet gets more and more power while climbing democratically the hierarchy levels. At level six it eventually reaches the majority to reach the presidency with certainty already at level ten. This is what is known in French as ``Un jeu de dupes".}
\label{abc1}
\end{figure}

Another illustration showing the evolution from  $p_{a,0}=0.48$,  $p_{b,0}=0.46$ and $p_{c,0}=0.06$ up to $p_{a,12}, p_{b,12}, p_{c,12}$ is exhibited in Figure (\ref{abc1}). Although the minority party C has only six percent of support at the bottom, its representative power will grow dramatically while climbing democratically the hierarchy levels taking advantage of the alliance set $(\alpha =0, \beta =0, \gamma =1 )$. At the seventh level it reaches the majority and then keeps growing to reach the presidency with certainty already at level 1 with $p_{a,10}=1$. This feature illustrates perfectly an agreement which indeed is  what French call ``Un jeu de dupes", or in English, "pulling the wool over your eyes". Such a situation implies a non uniform set of alliances as illustrated in the next Subsection.

\subsection{More on small party versus two large competing ones }

Therefore, it is not efficient for either one of the large parties to make a systematic alliance with the minority party C since it is equivalent to putting it in power after climbing up a sufficient number of levels. The aim of the agreement was to allow the corresponding large party to get the presidency. We thus need to revisit the alliance scheme. Consider now that parties A and C are still making an alliance, but in two steps. From the bottom and up a few levels, A votes for C in case of a (A B C) tie, thus boosting the C representation. But then, after for instance reaching the fifth level, starting from there, C have to vote for A in case of a tie. The precedent ``Un jeu de dupes" is now a balanced win-win agreement as seen using the same bottom supports $p_{a,0}=0.48, p_{b,0}=0.46, p_{c,0}=0.06$.

For the first five levels the new alliance scheme yields the same series as before with $p_{a,1}=0.47, p_{b,1}=0.44, p_{c,1}=0.09$, $p_{a,2}=0.45, p_{b,2}=0.41, p_{c,2}=0.14$, $p_{a,3}=0.43, p_{b,3}=0.37, p_{c,3}=0.20$, $p_{a,4}=0.40, p_{b,4}=0.31, p_{c,4}=0.29$, and $p_{a,5}=0.35, p_{b,5}=0.22, p_{c,5}=0.43$. This means an increase of $37\%$ for C and a $13\%$ decrease for A from the hierarchy bottom up to level 5.

\begin{figure}[ h]
\centering
\includegraphics[width=.5\textwidth]{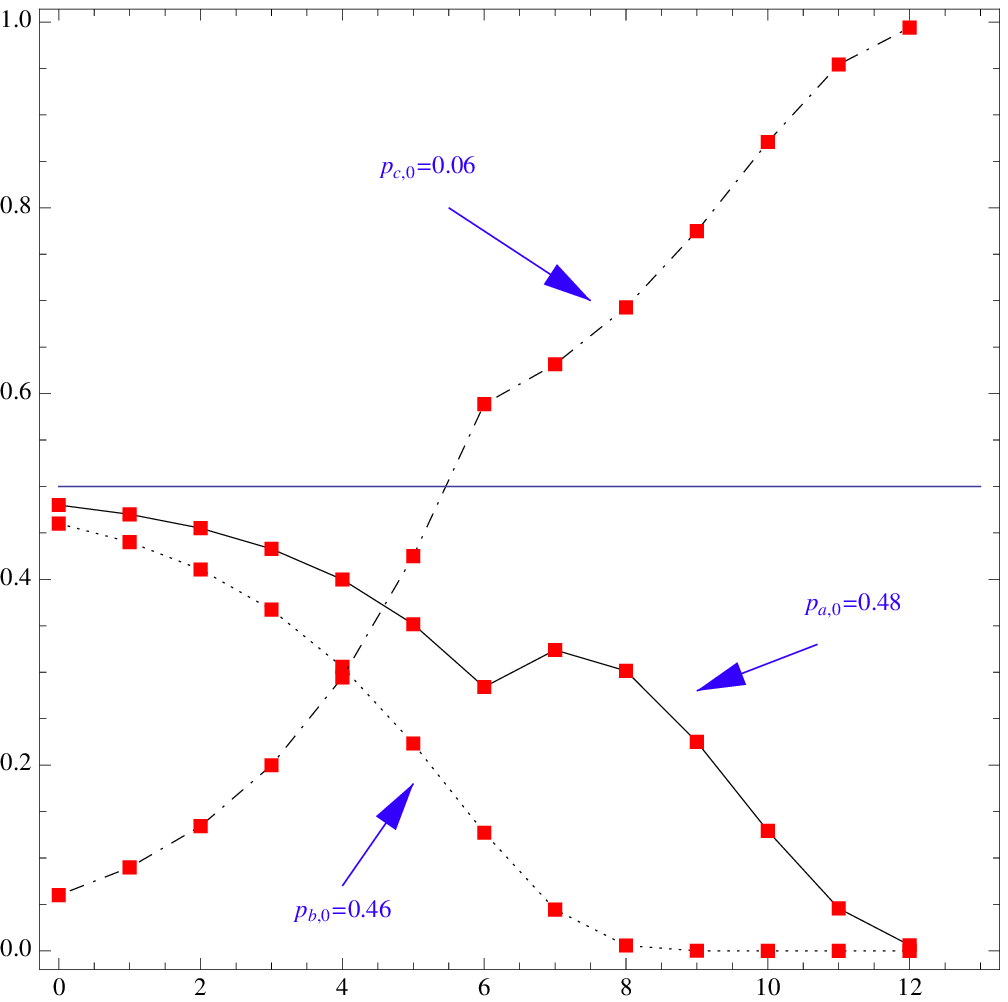}\hfill
\includegraphics[width=.5\textwidth]{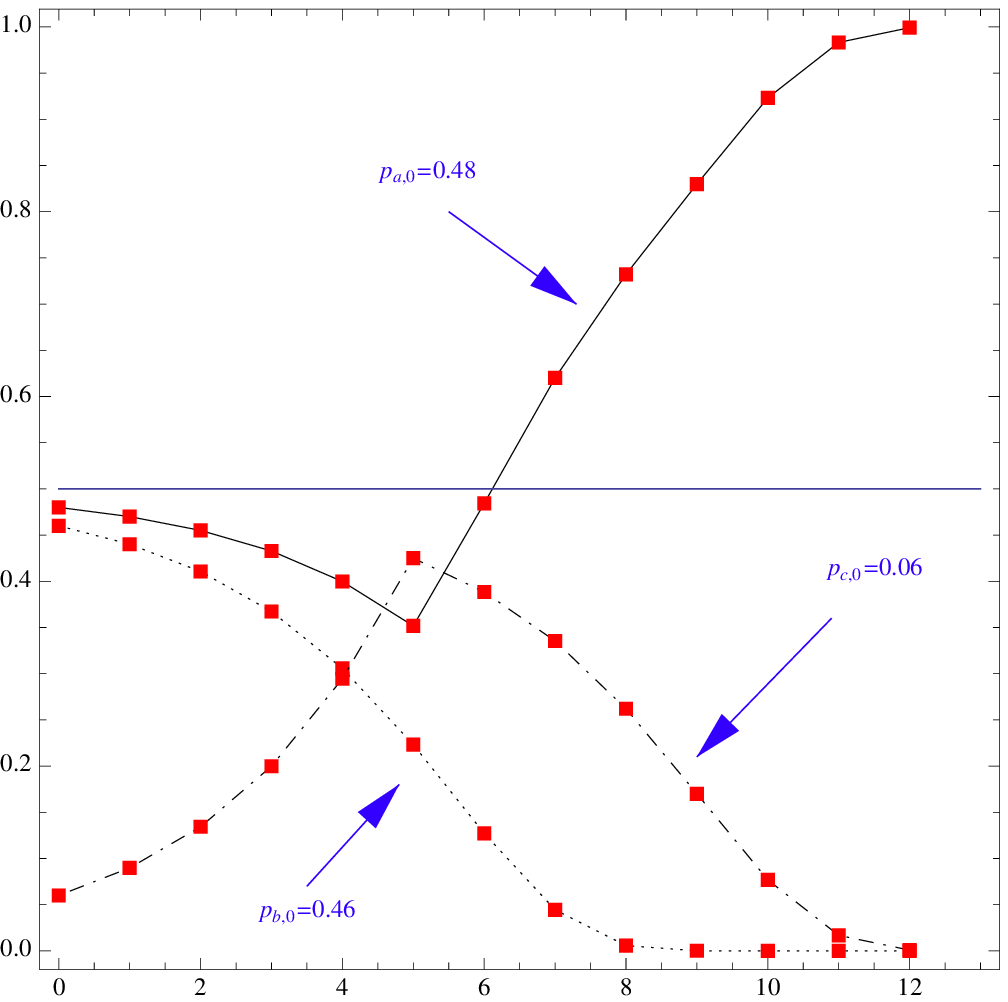}
\caption{The same as in Figure (\ref{abc1}) but here the set of alliance between A and C is different. Left: from the bottom up to level 5 not included, A votes for C in the case of a tie, but from level 5 and up party C votes for A in case of a tie. Now A eventually reaches the presidency although C gained a substantial increase in power. Right: same as left with a ``slight" difference. Here C votes for A at a tie only from level 6 up. The result is disastrous for A and great for C since at level 6 C has already gained the majority.}
\label{abc2}
\end{figure}

However, from level five and up while voting for the level above, in case of a (A B C) tie C votes for A. Now, after the first range of levels where A has lost support, the dynamics is reverse with A starting to grow steadily towards the presidency as shown in the left side of Figure (\ref{abc2}) with $p_{a,6}=0.48, p_{b,}=0.13, p_{c,6}=0.39$, $p_{a,7}=0.62, p_{b,7}=0.04, p_{c,7}=0.34$, $p_{a,8}=0.73, p_{b,8}=0.01, p_{c,8}=0.26$, $p_{a,9}=0.83, p_{b,9}=0.00, p_{c,9}=0.13$, $p_{a,10}=0.92, p_{b,10}=0.00, p_{c,10}=0.08$, $p_{a,11}=0.98, p_{b,11}=0.00, p_{c,11}=0.02$, $p_{a,12}=1, p_{b,12}=0.00, p_{c,12}=0.00$. 

The strategy reveals itself to be rather efficient for both A and C. It is worth stressing the fact that $p_{a,6}=0.48= p_{a,0}$ is a pure coincidence resulting from the rounding up. Indeed $p_{a,6}=0.484$ and could be quite different from $p_{a,0}$ depending on both the $p_{a,0}$ values.

The above two schemes of alliances demonstrates that the alliance must be dealt with great care. The right side of Figure (\ref{abc2}) shows the case of an error from A with a reverse in the vote alliance with C only from level six and up. It will be too late for A and C will eventually reach the presidency. A mistake at the level at which the alliance direction has to be switched is crucial. To avoid confusion, it is worth to stress that a voting agreement at a level $l$ yields an effect at level $(l+1)$.

For instance, within the above figures, selecting level six instead of five turns out to be a disaster for A as seen from the series with $p_{a,6}=0.28, p_{b,6}=0.13, p_{c,6}=0.59$; $p_{a,7}=0.32, p_{b,7}=0.04, p_{c,7}=0.64$; $p_{a,8}=0.30, p_{b,8}=0.01, p_{c,8}=0.69$; $p_{a,9}=0.22, p_{b,9}=0.00, p_{c,9}=0.78$; $p_{a,10}=0.13, p_{b,10}=0.00, p_{c,10}=0.87$; $p_{a,11}=0.05, p_{b,11}=0.00, p_{c,11}=0.95$; $p_{a,12}=0.00, p_{b,12}=0.00, p_{c,12}=1$; which are plotted in Figure (\ref{abc2}). 

The reversing voting bias came in too late to make A win since at level 6 C has already gained the majority making the dynamics leading towards the C attractor independently of the alliance set. The result at level 7 with $p_{a,7}=0.32>p_{a,6}=0.28$ can be misleading since afterwards we have  $p_{a,8}<p_{a,6}$ and so forth towards zero. As soon as a party gets the majority it is certain to win the presidency provided there exists a sufficient number of levels. Accordingly any agreement from one of the big party should prevent totally the small party from reaching ever the majority at any level.

\subsection{A large party just below $50\%$ versus two equal medium size ones}

\begin{figure}[ h]
\centering
\includegraphics[width=.5\textwidth]{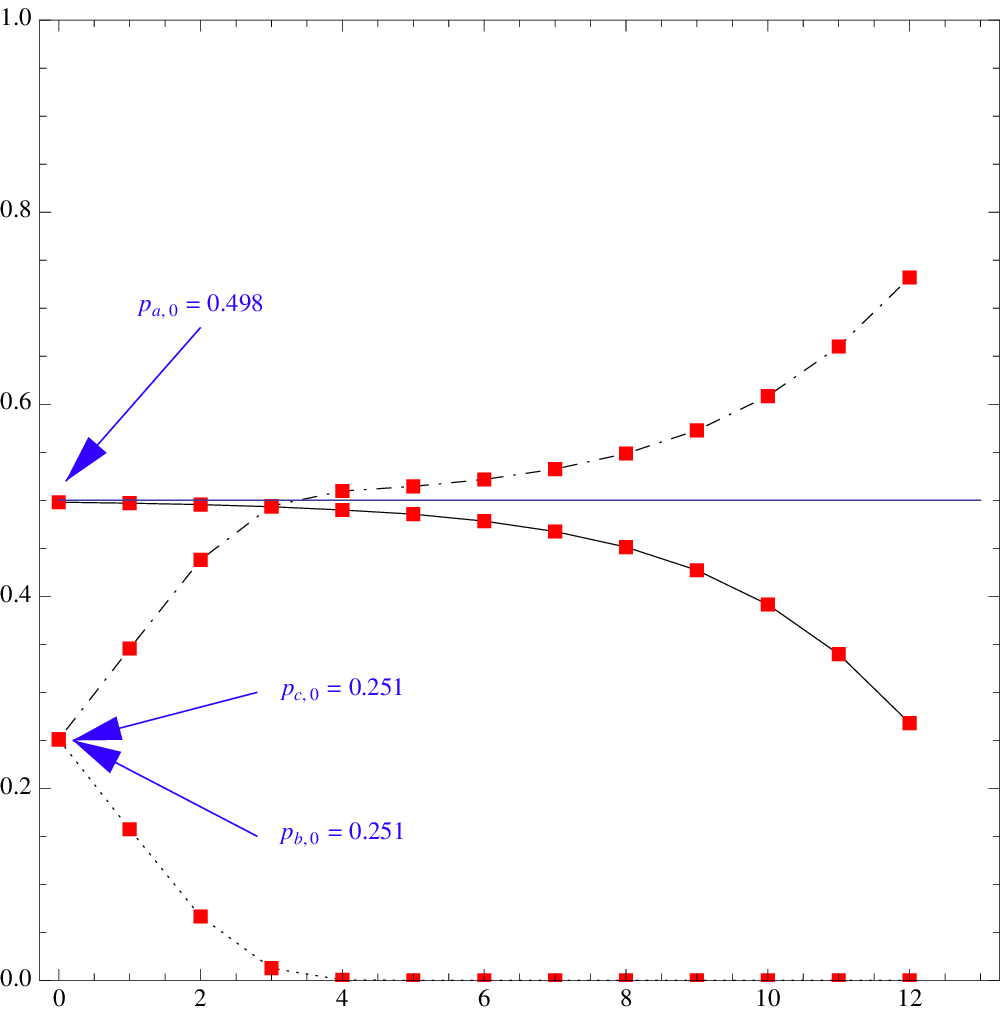}\hfill
\includegraphics[width=.5\textwidth]{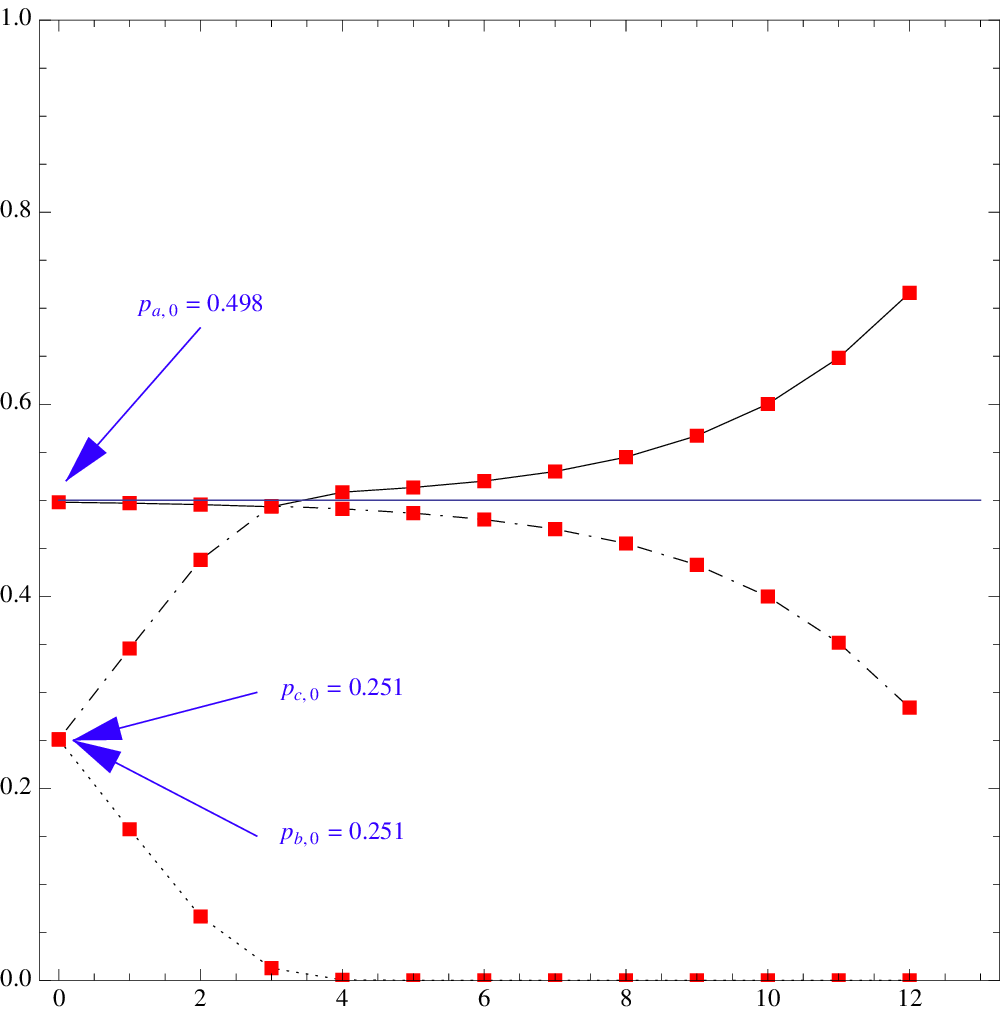}
\caption{ The respective variation of power sharing for A, B, C from a bottom support $p_{a,0}=0.498$ (solid line), $p_{b,0}=0.251$ (dotted line), $p_{c,0}=0.251$ (dashed line) as a function of the hierarchy level up to level 12. Left: from the bottom up to level 4 included, A votes for C in the case of a tie, but from level 5 and up party C votes for A in case of a tie.  Right: same as left with a ``slight" difference. Here C votes for A at a tie already from level 4 up.}
\label{aa}
\end{figure}

After investigating the case of two big parties versus a small one it is worth to look briefly at the case of one big party just below the winning level of $50\%$ competing with two opposite medium size parties. Precisely, Figure (\ref{aa}) shows the case $p_{a,0}=0.498, p_{b,0}= p_{c,0}=0.251$ with an agreement between the big party A and party C in favor of C up to the first 4 levels and in favor of A for the next  8 levels of a 12 level hierarchy (left part of the Figure). 

While B disappears totally from level 4 and up, party A loses the lead at the benefit of party C, which turns to majority from level 4 and up. At the top hierarchy the probability for C to get the presidency is 0.73 while it is only 0.34 for A although it has almost fifty at the bottom with $p_{a,12}=0.27, p_{b,12}= 0, p_{c,12}=0.73$.

A slightly different agreement placing the shift in favor of A from level 3 instead of 4 (right part of the Figure) modifies substantially the outcome. While C is still getting more representatives even getting ahead of A at level 3 with $p_{a,3}=0.493, p_{b,3}= 0.013, p_{c,3}=0.494$, the dynamics is reversed at the higher levels with at the top level 
$p_{a,12}=0.72, p_{b,12}= 0, p_{c,12}=0.28$.

The case illustrates once more the very subtle sensibility to where to place the shit to favor one party against the other one within a coalition agreement.

\subsection{ Illustration for different sets of $(\alpha, \beta, \gamma)$}

\begin{figure}[ htpb ]
\centering
\includegraphics[width=.5\textwidth]{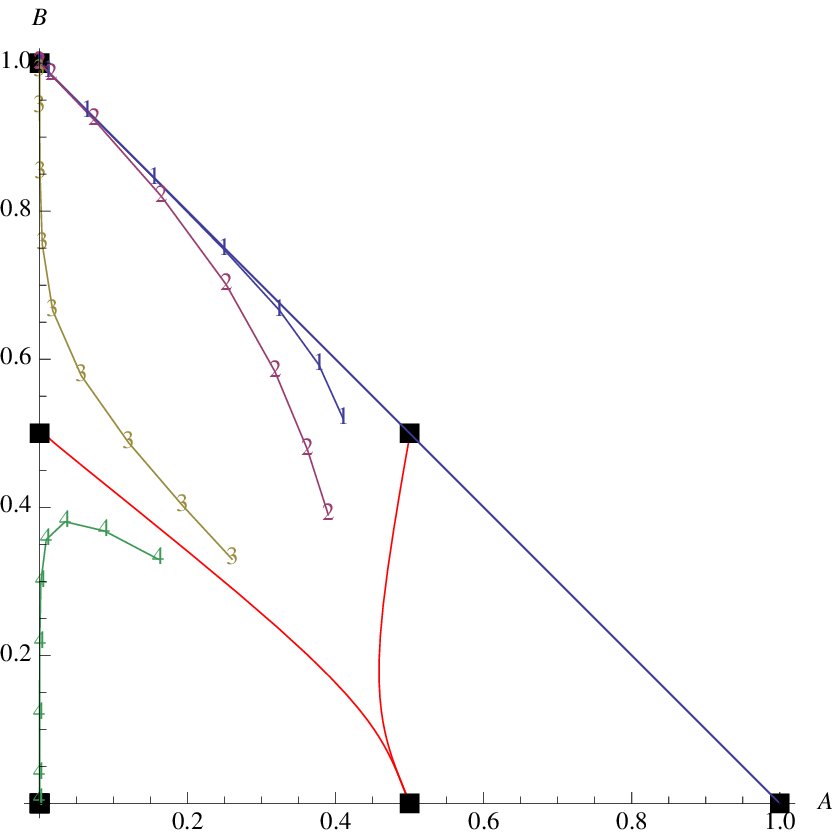}\hfill
\includegraphics[width=.5\textwidth]{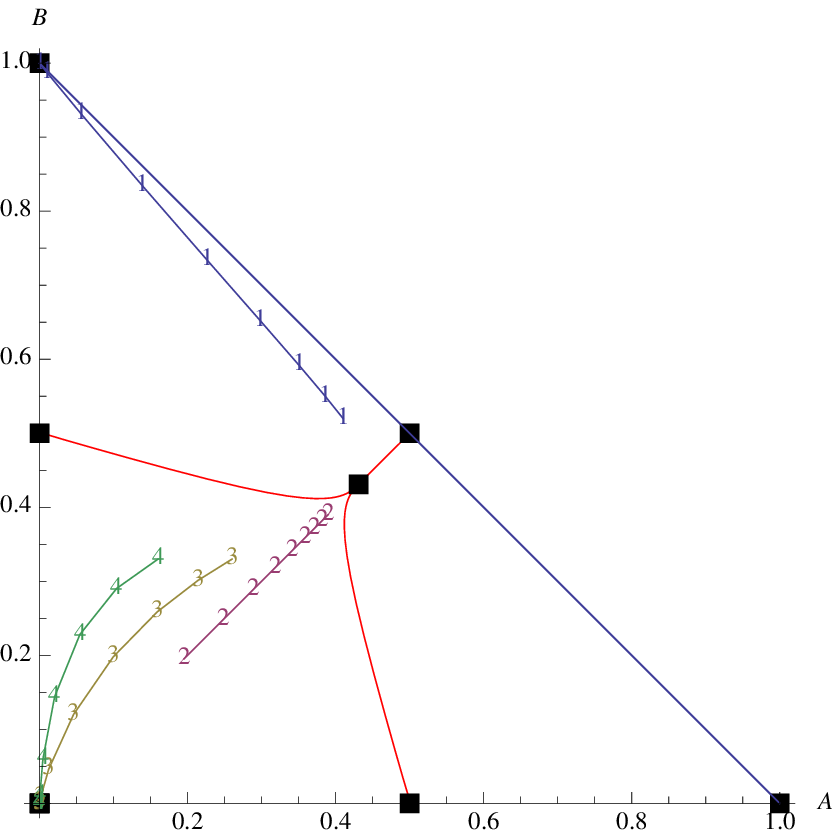}\hfill
\includegraphics[width=.5\textwidth]{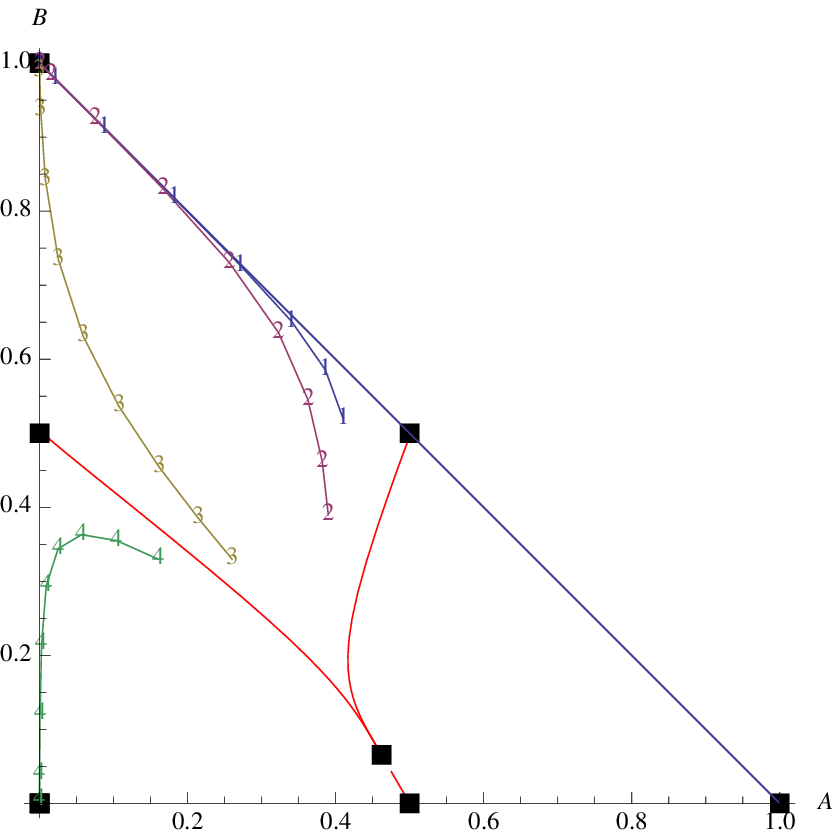}\hfill
\includegraphics[width=.5\textwidth]{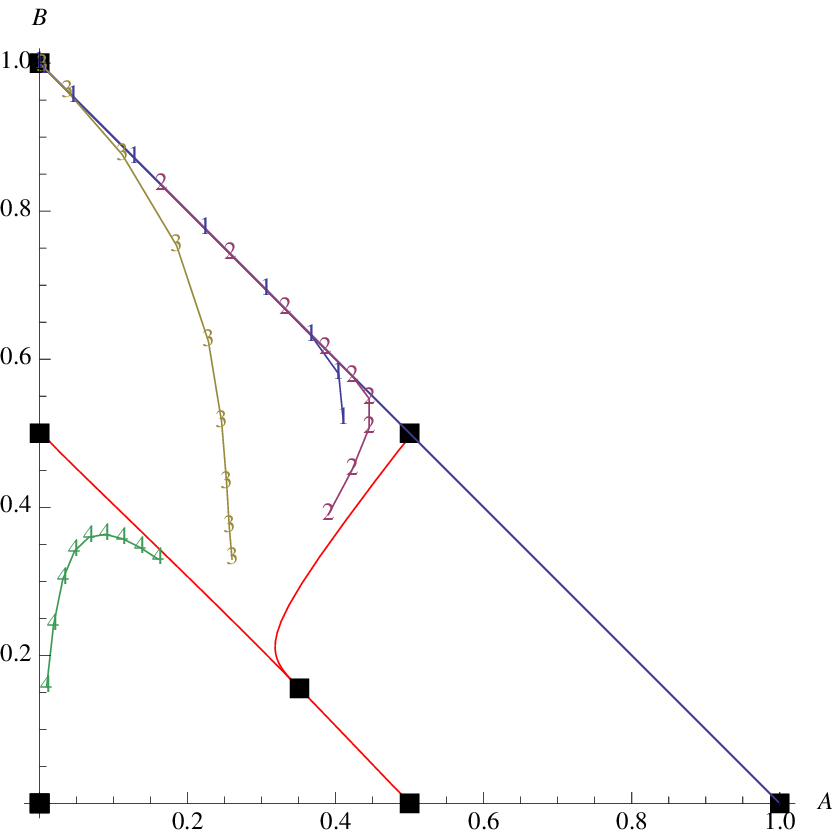}\hfill
\caption{Voting flow for eight hierarchical levels from the four bottom values $B_1: (0.41, 0.52, 0.07)$ (noted with ``1");  $B_2: (0.39, 0.39, 0.22)$ (noted with ``2"); $B_3: (0.26, 0.33, 0.41)$ (noted with ``3");  $B_4: (0.16, 0.33, 0.51)$ (noted with ``4"); with the four set of alliances $A_1: (0.12, 0.70, 0.18)$; $A_2: (0.22, 0.22, 0.56)$; $A_3: (0.22, 0.62, 0.16)$; $A_4: (0.42, 0.56, 0.02)$. The black squares are the various fixed points with the triangle vertices being the three attractors. When an attractor is not reached the presidency winner is probabilistic.}
\label{g1} 
\end{figure}

\begin{figure}[ htpb ]
\centering
\includegraphics[width=.5\textwidth]{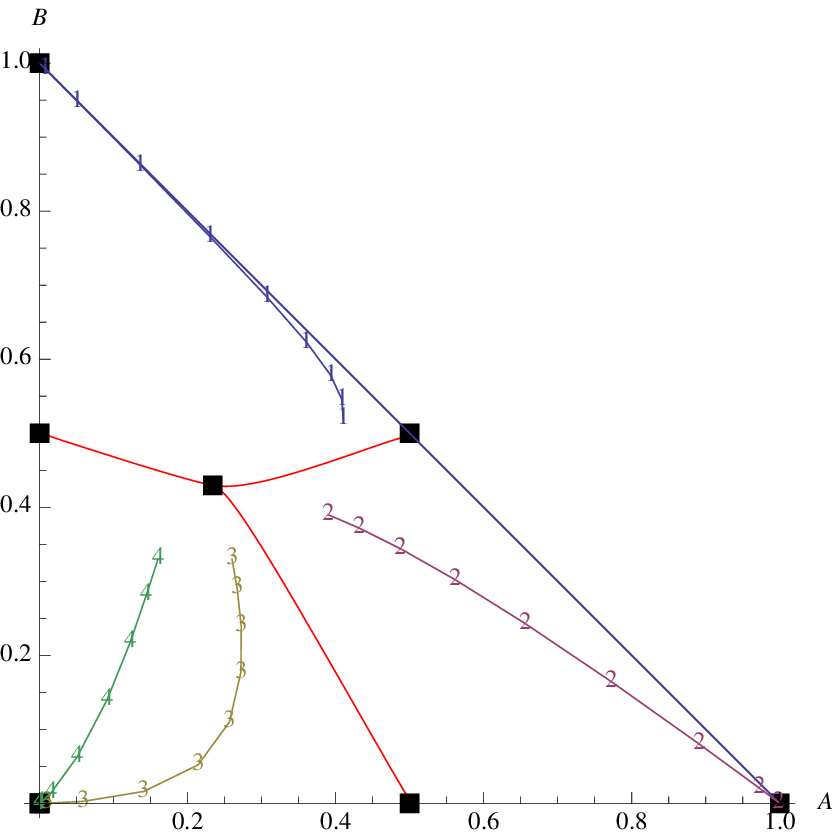}\hfill
\includegraphics[width=.5\textwidth]{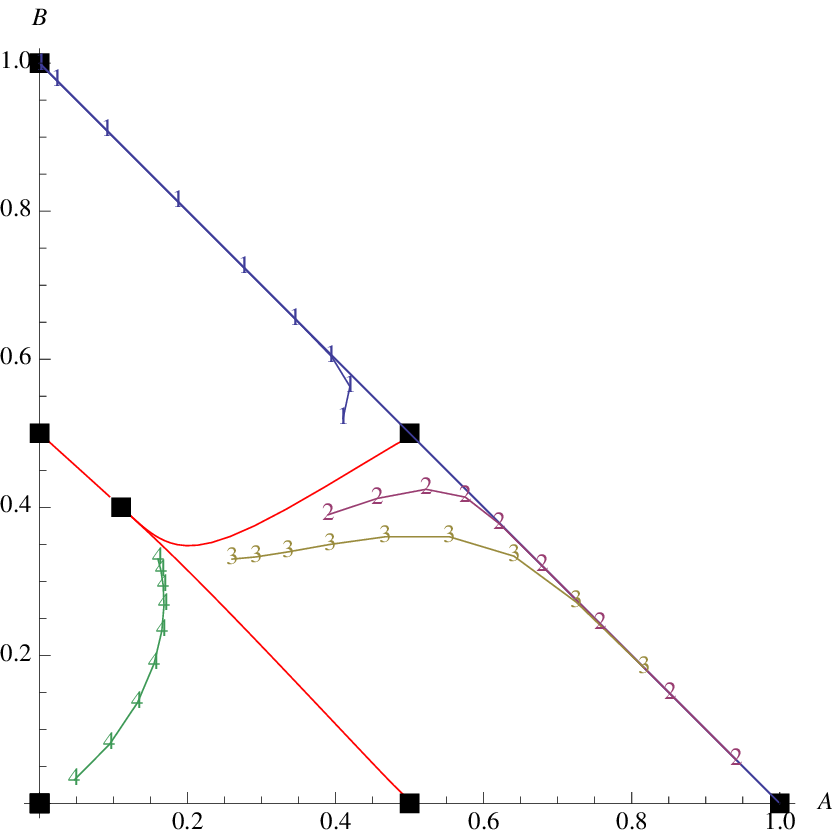}\hfill
\includegraphics[width=.5\textwidth]{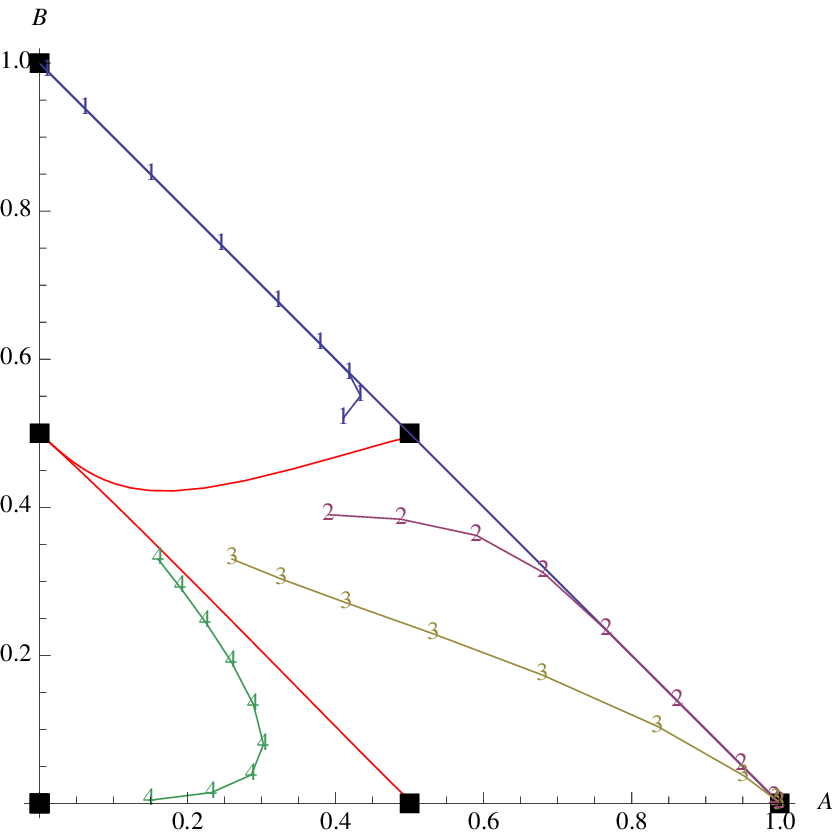}\hfill
\includegraphics[width=.5\textwidth]{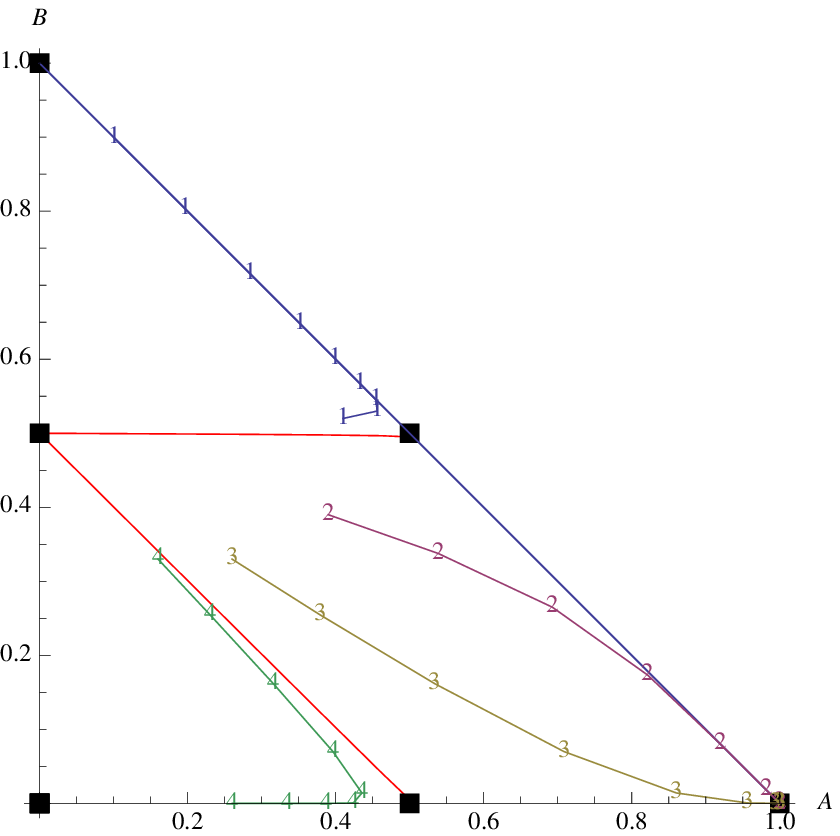}\hfill
\caption{Voting flow for eight hierarchical levels from the four bottom values $B_1: (0.41, 0.52, 0.07)$ (noted with ``1");  $B_2: (0.39, 0.39, 0.22)$ (noted with ``2"); $B_3: (0.26, 0.33, 0.41)$ (noted with ``3");  $B_4: (0.16, 0.33, 0.51)$ (noted with ``4"); with the four set of alliances $A_5: (0.47, 0.17, 0.36$); $A_6: (0.59, 0.37, 0.04)$; $A_7: (0.75, 0.23, 0.02)$; $A_8: (1, 0, 0)$. The black squares are the various fixed points with the triangle vertices being the three attractors. When an attractor is not reached the presidency winner is probabilistic.}
\label{g2}
\end{figure}

To enhance the associated various voting dynamics we show two Figures (\ref{g1}, \ref{g2}) with a series of snapshots combining four initial (bottom) probability sets $B_i$  for $(p_{a,0}, p_{b,0}, p_{c,0})$ with $i=1, 2, 3,  4$ with eight alliance probabilities $A_j$ for $(\alpha, \beta, \gamma)$ where $j=1,2,..., 8$. These four bottom cases are: \\ \\
$B_1: (0.41, 0.52, 0.07)$ (noted with ``1");  \\
$B_2: (0.39, 0.39, 0.22)$ (noted with ``2");  \\
$B_3: (0.26, 0.33, 0.41)$ (noted with ``3");  \\
$B_4: (0.16, 0.33, 0.51)$ (noted with ``4");  \\ \\
with the eight alliance probabilities  \\ \\
$A_1: (0.12, 0.70, 0.18)$; \\
$A_2: (0.22, 0.22, 0.56)$; \\
$A_3: (0.22, 0.62, 0.16)$; \\
$A_4: (0.42, 0.56, 0.02)$; \\ 
$A_5: (0.47, 0.17, 0.36$); \\
$A_6: (0.59, 0.37, 0.04)$;  \\
$A_7: (0.75, 0.23, 0.02)$;  \\
$A_8: (1, 0, 0)$. \\

The associated 32 combinations are shown in Figs. (\ref{g1}) and (\ref{g2}) with four cases in each of the four figure parts, all for eight hierarchical levels. After eight levels when an attractor is not reached the presidency winner is probabilistic. In those cases more levels would be necessary to make the presidency outcome deterministic. Non integer values for $(\alpha, \beta, \gamma)$ represent averages over different voting groups at all levels. In particular it makes it possible to take into account local defections with respect to global party agreements.

In Figure (\ref{g1}) we have $A_1$ in upper left; $A_2$ in upper right; $A_3$ in lower left;   $A_1$ in lower right; and In Figure (\ref{g2}) we have $A_5$ in upper left; $A_6$ in upper right; $A_7$ in lower left;   $A_8$ in lower right. The series of Figures (\ref{g1}, \ref{g2}) represent the two dimensional projection OAB of the plane ABC as shown in Figure (\ref{fixed-pt3}). The two frontiers which determine the respective basins of attraction of each pure attractor are calculated numerically and drawn on the Figures. The origin O is the projection of the C attractor. All fixed points are also included.

\section{Conclusion}

Above illustrations demonstrate the drastic effect of slight differences in local alliances to control democratically a bottom-up hierarchy.  To design a winning strategy requires to have access to the frontiers of the dynamics flow, which are determined the faith of the elections. Their respective location depends on the  effective average value of $( \alpha, \beta, \gamma )$. What matters is the positioning within the ``golden triangle" determined by the simultaneous tree constraints $p_{a,0}<\frac{1}{2}$ and $p_{b,0}<\frac{1}{2}$, and $p_{c,0}<\frac{1}{2}$, to win the presidency. On this basis it is worth stressing that the distribution of the area of this golden triangle in favor of each one of the party depends solely on  $( \alpha, \beta, \gamma )$ independently of the values of $p_{a,0}, p_{b,0}, p_{c,0}$. Those average values are a function of  the alliances set up by the parties as well as the discipline to apply these party agreements by local representatives. Once the frontiers are set, the initial values determines the flow direction. 

The voting flows shown in Figures (\ref{g1}, \ref{g2}) are self explicative and do not require additional comments. It puts a new quantitative light on the dramatic effects of setting up alliances together with their local reliability that needs to be implemented. 

Three-party competition exhibits a rather rich and subtle dynamics driven by the overall averages of local and global alliances the various parties can build. Slight, a priori insignificant differences may induce drastic and counter intuitive changes.

To conclude a word of caution is in order. The use of above results for applications in real hierarchical organizations can turn anti-democratic. However at the same time, the knowledge of those results can help improving the democratic functioning of non democratic structures. It is fortunate that Machiavel was missing the knowledge of renormalization group.


\end{document}